\documentclass{aa}
\pdfoutput=1
\usepackage[english]{babel}
\usepackage[utf8]{inputenc}
\usepackage[varg]{txfonts}
\usepackage{natbib,twoopt}
\usepackage{graphicx}
\usepackage{amsmath}
\usepackage[breaklinks=true]{hyperref} 
\bibpunct{(}{)}{;}{a}{}{,} 
\makeatletter
\newcommandtwoopt{\citeads}[3][][]{\href{http://adsabs.harvard.edu/abs/#3}%
{\def\hyper@linkstart##1##2{}%
\let\hyper@linkend\@empty\citealp[#1][#2]{#3}}}
\newcommandtwoopt{\citepads}[3][][]{\href{http://adsabs.harvard.edu/abs/#3}%
{\def\hyper@linkstart##1##2{}%
\let\hyper@linkend\@empty\citep[#1][#2]{#3}}}
\newcommandtwoopt{\citetads}[3][][]{\href{http://adsabs.harvard.edu/abs/#3}%
{\def\hyper@linkstart##1##2{}%
\let\hyper@linkend\@empty\citet[#1][#2]{#3}}}
\newcommandtwoopt{\citeyearads}[3][][]%
{\href{http://adsabs.harvard.edu/abs/#3}
{\def\hyper@linkstart##1##2{}%
\let\hyper@linkend\@empty\citeyear[#1][#2]{#3}}}
\makeatother

\title{The environment of radio sources in the VLA-COSMOS Survey field}

\author{
N. Malavasi \inst{1}\thanks{\textit{E-mail contact:} \href{mailto:nicola.malavasi@unibo.it}{nicola.malavasi@unibo.it}} 
\and 
S. Bardelli \inst{2} 
\and 
P. Ciliegi \inst{2} 
\and 
O. Ilbert \inst{3} 
\and 
L. Pozzetti \inst{2} 
\and 
E. Zucca \inst{2}}

\institute{
University of Bologna, Department of Physics and Astronomy (DIFA), v.le Berti Pichat 6/2 - 40127 Bologna, Italy 
\and 
INAF--Osservatorio Astronomico di Bologna, via Ranzani 1 - 40127 Bologna, Italy 
\and 
Aix Marseille Universit\'{e}, CNRS, LAM (Laboratoire d'Astrophysique de Marseille), UMR 7326, 13388 Marseille, France}

\date{Received October 14, 2014 / Accepted December 24, 2014}

\abstract
{Several problems regarding the process of galaxy formation are nowadays still open. One of them is the role played by Active Galactic Nuclei (AGN) phenomena in contributing to galaxy build-up and in particular to Star Formation (SF) quenching. On the other hand, the theory of AGN formation predicts these phenomena to be correlated with the host-galaxy environment, therefore opening for links between SF quenching, environment and AGN phenomena in the galaxy formation and evolution paradigm.}
{This work is focused on the study of the correlation between environmental density, radio AGN presence and the probability of a galaxy to host a radio Active Galactic Nucleus.}
{Using data from the photometric COSMOS survey and its radio 1.4 GHz follow-up (VLA-COSMOS), a sample of radio AGNs has been defined. The environment has been studied throughout the use of the richness distributions inside a parallelepipedon with base side of 1 Mpc and height proportional to the photometric redshift precision. Richness distributions have been compared as a function of both the redshift and the relative evolution of the stellar masses of galaxies and AGN hosts up to $z = 2$.}
{Radio AGNs are always located in environments that are significantly richer and denser than those around galaxies in which radio emission is absent. Therefore, the environment seems to enhance the probability of a galaxy to host a radio AGN. Moreover, a distinction between high power and low power radio AGNs leads to the conclusion that the significance in the environmental effect is maintained only for low power radio sources. By studying the evolution of stellar masses it is possible to conclude that radio AGN presence is a phenomenon related to quiescent galaxies up to $z = 2$, with a significant increase of the fraction of quiescent galaxies hosting a radio AGN with decreasing redshift. Hints of an environmental effect are present as well.}
{The results found with this work lead to conclude that denser environments play a significant role in enhancing the probability of a galaxy to host a radio AGN and in particular low power ones.}
\keywords{Galaxies: active - Galaxies: formation - Galaxies: evolution - Galaxies: mass function - Galaxies: star formation - Galaxies: clusters: general}

\begin{document}

\maketitle

\section{Introduction}
\label{intro}
It is well known that the morphology of galaxies is related with the local density in clusters \citepads{1980ApJ...236..351D}. Moreover, luminosity functions \citepads{2009A&A...508.1217Z}, colours \citepads{2010A&A...524A...2C} and stellar masses \citepads{2010A&A...524A..76B} show dependencies up to redshift $z \sim 1$ with the density field. Nevertheless, the mechanism that transforms star-forming galaxies into quiescent ones in dense environments is not yet well understood. Recently, \citetads{2010ApJ...721..193P} suggested a model where the star formation quenching is separable in a mechanism depending on the galaxy mass and on the density; the first mechanism acting principally on massive objects at the centre of halos, while the second regarding satellite galaxies.

The mass quenching can be related to the so called \textquotedblleft feedback\textquotedblright. This term simply indicates an energy input which causes an early stop of the star formation in a galaxy. This energy input was first introduced to reconcile cosmological simulations with observations \citepads[see for example][]{2004ApJ...600..580G, 2006MNRAS.365...11C} and since then it has been naturally associated with the activity of an Active Galactic Nucleus (AGN).

By noting that X-ray and radio AGNs have different clustering properties, \citetads{2009ApJ...696..891H} suggested that the central AGN co-evolves with the host galaxy: while the host galaxy transforms from a star-forming to a quiescent one, the AGN passes from a quasar, X-ray emitter phase to a radio-galaxy one. These transformations happen at earlier epochs for halos of higher mass. This evolution of AGN type with galaxy transformation could be the reason behind the estimate that radio AGNs constitute $\sim 10\%$ of the whole Active Galactic Nuclei population \citepads[see for example][]{2009ApJ...696..891H}, although projection effects and the limiting flux of current radio surveys may also play a role. The motivation of such a small fraction of radio to total AGNs is still not fully understood. Since the analysis conducted in this work focuses only on radio AGN samples, in further text it will be explicitly referred to this kind of sources.

In the scenario developed by \citetads{2009ApJ...696..891H} and \citetads{2010ApJ...721..193P}, the environment is still related to the star formation quenching of massive galaxies in an indirect way, in the sense that massive galaxies (where the mass quenching is effective) reside preferentially in high density environments, where at low redshifts early-type galaxies dominate \citepads{2012ApJ...744...88Q,2011MNRAS.413.1678C}. Moreover, it was already known that many radio AGNs reside in early-type galaxies \citepads{1996AJ....112....9L} and that the probability for a galaxy to host an AGN is increasing with stellar mass \citepads{2009A&A...495..431B}. Also, \citetads{2010A&A...511A...1B}, on the basis of the zCOSMOS 10k sample \citepads{2007ApJS..172...70L, 2009ApJS..184..218L} showed that 
the fraction  of radio active early-type galaxies is an increasing function of local density.

However, what it is not clear is what is meant when environment is considered, since its definition is related to the spatial scale studied. In fact, environments estimated with nearest neighbour methods consider different scales at different redshifts in a non trivial way. For example, at least at low redshifts, \citetads{2004MNRAS.351...70B} claims that the larger scale (\textit{i.e.} the scale typical of groups and clusters) is more important than the smaller scale (typical of galaxy pairs or companions) to determine the AGN radio emission.
 
In this work the environment of radio sources of the VLA-COSMOS survey \citepads{2007ApJS..172...46S}, cross-identified with the COSMOS photometric redshift sample \citepads{2009ApJ...690.1236I}, is explored. With respect to the \citetads{2009A&A...495..431B} work, this sample allows an increase in both redshift range and statistics, but at the price of having a larger error in the redshifts (photometric \textit{versus} spectroscopic redshifts), which excludes the possibility of a small spatial scale definition for the environment. The scale of the environment is fixed at 1 Mpc, in order not to depend on the scale-redshift relation.

The data and the method used to define the radio AGN sample are described in sections \ref{data} and \ref{agndefinition} respectively. In section \ref{method} the method to compare the overdensity richness distributions is presented and in sections \ref{results} and \ref{agnenvironment} the results for various sub-samples are discussed. In section \ref{correlationclusters} a correlation with known clusters and groups catalogues is attempted and in section \ref{massfunctions} the integrated mass functions of radio AGNs and normal galaxies are presented and analysed. Conclusions are drawn in section \ref{conclusions}. Throughout the text, masses are expressed in units of Solar masses, while radio luminosities are expressed in SI units. The assumed cosmology is $\Omega_{\varLambda}=0.7$, $\Omega_{m}=0.3$ and $H_0=70$ km s$^{-1}$ Mpc$^{-1}$. For the SED fitting procedure, in order to derive stellar masses and Star Formation Rates, \citetads{2010ApJ...709..644I} used SED templates produced 
with a \citetads{2003PASP..115..763C} IMF.

\section{Data}
\label{data}
In this work, two main datasets have been used. Optical data (with photometry in 30 bands from UV to mid-IR) have been used to estimate photometric redshifts for all the sources, to determine the environment around every object in every sample and to extract \textit{Control} samples as explained later in the text.

Radio data at 1.4 GHz, instead, have been used to create the radio source and AGN samples, around which the environment has been studied.

\subsection{Optical data}
\label{opticaldata}
The data at optical wavelengths are constituted of the photometric sample of \citetads{2007ApJS..172...99C}, used in the construction of the version 1.8 of the photometric redshift catalogue of \citetads{2009ApJ...690.1236I}.
This sample is a compilation of photometric data taken from the COSMOS survey \citepads{2007ApJS..172....1S}, which covers a field of an area of about 
$1.4^{\degr}\times1.4^{\degr}$ and is centred at R.A. (J2000.0) =
$10^h00^m28.6^s$ and Dec (J2000.0) = $+02^{\degr}12\arcmin21.0\arcsec$.

In \citetads{2009ApJ...690.1236I}, photometric redshifts have been calculated through a Spectral Energy Distribution (SED) fitting procedure, using 30 broad-, 
intermediate- and narrow-band filters from UV to mid-IR frequencies 
(see their Table 1). The fit has been performed with a $\chi^2$ minimization algorithm on a template library using the \textit{Le Phare} code (S. Arnouts \& O. Ilbert). 

The redshifts obtained with the described procedure have been compared with those obtained using the zCOSMOS spectroscopic survey \citepads{2007ApJS..172...70L}.

The accuracy of the photometric redshifts ($z_p$) is estimated to be
\begin{equation}
\sigma_{\Delta z/(1+z)}  = 0.06 
\end{equation}

This value will be assumed throughout this work. The error has been conservatively selected as the maximum value of those obtained as a function of both magnitude and redshift in \citetads[][see their Figure 9]{2009ApJ...690.1236I}, taking into account the depth of the optical catalogue used for the analysis. In order to check the consistency of such a choice, the effect of degrading the value to $\sigma_{\Delta z/(1+z)}  = 0.1 \div 0.2$ for high redshift sources has been tested. As explained further in the text, it has been found that it does not change significantly the results of this work.

Other sources of uncertainties in the photometric redshifts determination which must be kept in mind are the so called \textit{catastrophic errors}. The definition of {\it catastrophic errors} is applied to those sources for which the redshift calculation fails in the form of
\begin{equation}
\frac{|z_p - z_s|}{1+z_s} > 0.15  
\end{equation}
where $z_s$ is the spectroscopic redshift. It is estimated that the fraction $\eta$ of catastrophic errors rises from $1\%$ to $20\%$ going from $i^+ < 22.5$ to $i^+ \sim 24$.

Only objects that are not in masked areas and with magnitude $i^+ < 26.5$ are considered. This magnitude limit approximately corresponds to the point where the magnitude-number counts histogram stops increasing. At this limit the average number of filters used for the photometric redshifts determination is 18. Although the effect of such a deep cut in limiting magnitude is to increase the uncertainties in the physical quantities derived from the optical photometry, it has also the advantage of greatly increasing the statistics for the environment estimate. For this reason, this cut will be assumed for all samples in this work. As explained extensively in section \ref{method}, tests have been performed by introducing brighter magnitude cuts, from $i^+ < 25.5$ to $i^+ < 24$, to check the problematics related to including in the analysis also very faint objects. These tests showed that the magnitude cut has no major effects on the results found with this work.

From the optical catalogue a sample of  823\,939 sources with optical data and measured photometric redshift has been extracted (hereafter \textquotedblleft O sample\textquotedblright). Star Formation Rates (SFR) and stellar masses ($M^{\ast}$) are both derived through SED fitting with population synthesis models \citepads[see][]{2010ApJ...709..644I}, together with other physical parameters.

\subsection{Radio Data}
\label{dataradio}
The radio data are taken from the VLA-COSMOS Large Project version 2.0 catalogue \citepads{2007ApJS..172...46S, 2010ApJS..188..384S}, whose observations have been carried out with the Very Large Array in its A configuration. This grants a resolution of about $1.5\arcsec$ at Full Width Half Maximum (FWHM) in the observation wavelength of 1.4 GHz. A total of 23 pointings was necessary to cover the full COSMOS field, for 240 hours of observation, performed between 2004 and 2005. The observations reached a sensitivity  of about 11 $\mu$Jy r.m.s. and the $5 \sigma$ detection limit catalogue contains 2417 radio sources, 78 of which have been classified as multi-component radio sources.

The search for the optical counterparts has been done using the photometric redshift sample (version 1.8) by \citetads{2009ApJ...690.1236I}, without limitation on the optical magnitude, while for the radio catalogue the VLA-COSMOS Large Project version 2.0 has been used, but without the 78 multi-component radio sources (2339 radio sources).

The method used for the optical identification is the likelihood ratio method, described in \citetads{1992MNRAS.259..413S}, \citetads{2005A&A...441..879C} and \citetads{2007ApJS..172..353B}. For the likelihood ratio analysis it has been adopted a Q value (the probability that the optical counterpart of the radio sources is brighter than the magnitude limit of the optical catalogue) of 0.9 and a likelihood ratio cut-off of 0.1 ($1-Q$).  Using a likelihood ratio cut-off equal to $1-Q$ ensures that all the optical counterparts of the radio sources with only one identification and a likelihood ratio greater than the cut-off value have a reliability greater than 0.5.  With these thresholds 2069 radio sources with an optical counterpart have been found, $\sim 26$ of which ($\sim 1.3$\%) could be spurious positional coincidences. Applying to the 2069 identified radio sources the same cuts used to define the O sample (objects not in masked areas and with magnitude $i^+ < 26.5$) a sample of 1806 optically identified 
radio souces has been created (hereafter \textquotedblleft R sample\textquotedblright). The data analysis has, nevertheless, been performed only out to $z \le 2.0$, \textit{i.e.} on a selection from the R sample of 1427 objects. The sizes of the various samples can be found in Table \ref{redshiftbins}, while a summary of the sample definitions can be found in Table \ref{samples}.

\begin{table}
\caption{Size of every sample and redshift range of every bin used in the data analysis.}
\centering
\begin{tabular}{c c c c c}
\hline\hline
Bin & $z$ & R sample & MR sample & AGN sample  \\
\hline
Bin 1 & 0 - 0.7 & 532 & 425 & 119  \\
Bin 2 & 0.7 - 1 & 320 & 290 & 100  \\
Bin 3 & 1 - 2   & 575 & 503 &  53  \\
\hline
\end{tabular}
\label{redshiftbins}
\end{table}

\begin{table*}
\caption{Samples definition.}
\centering
\begin{tabular}{c c}
\hline\hline
Sample Name & Definition \\
\hline
O   & Optical objects, $i^+ \le 26.5$ \\
R   & Optically identified radio objects, $i^+ \le 26.5$ \\
MR  & Optically identified radio objects, $i^+ \le 26.5$, $\log(M^{\ast}) \ge 10$ \\
AGN & AGNs, $i^+ \le 26.5$, $\log(M^{\ast}) \ge 10$, $\log(\frac{SSFR}{yr^{-1}}) \le -11$\\
RO  & Control sample of R, extracted from O \\
MO  & Control sample of MR, extracted from O, same $M^{\ast}$ distribution \\
QO  & Control sample of AGN, extracted from O, same $M^{\ast}$ distribution, same SSFR range \\
\hline
\end{tabular}
\label{samples}
\end{table*}

\begin{figure}
\resizebox{\hsize}{!}{\includegraphics[angle=90]{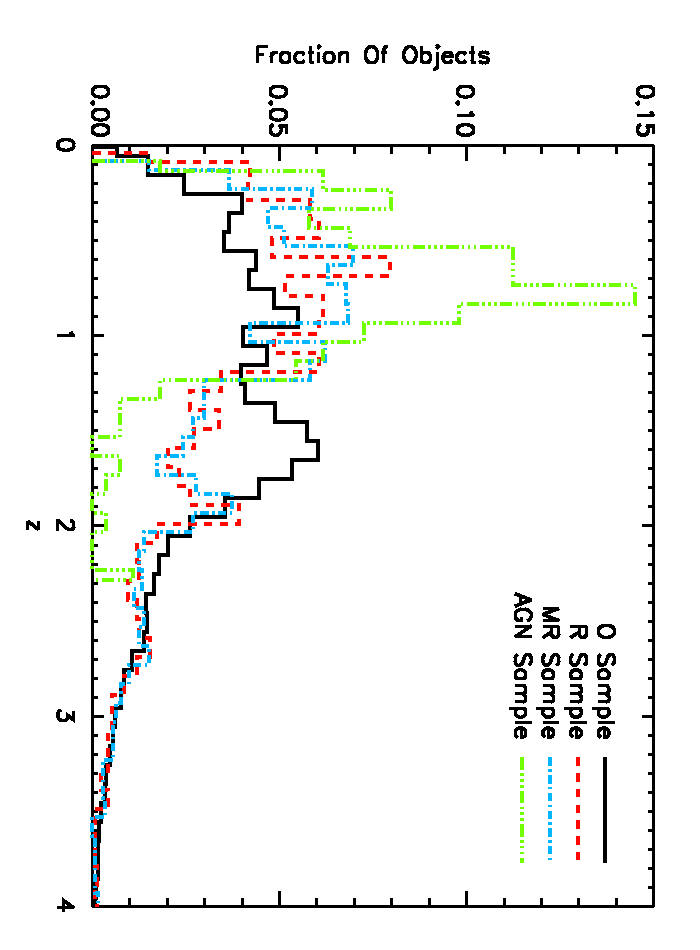}}
\caption{\textit{Redshift Distributions}. The black solid line refers to the O sample, the red dashed one to the R sample, the cyan dot-dashed line refers to the MR sample and the green triple dot-dashed one refers to the AGN sample.}
\label{zdist}
\end{figure}

In Figure \ref{zdist} the redshift distributions for the O, R, MR and AGN samples (the last two will be introduced further on in the text) are compared. It can be seen that the distributions referring to radio galaxies peak at lower redshift, while the distribution for the optical galaxies is wider in a more extended redshift range.

\begin{figure}
\resizebox{\hsize}{!}{\includegraphics[angle=90]{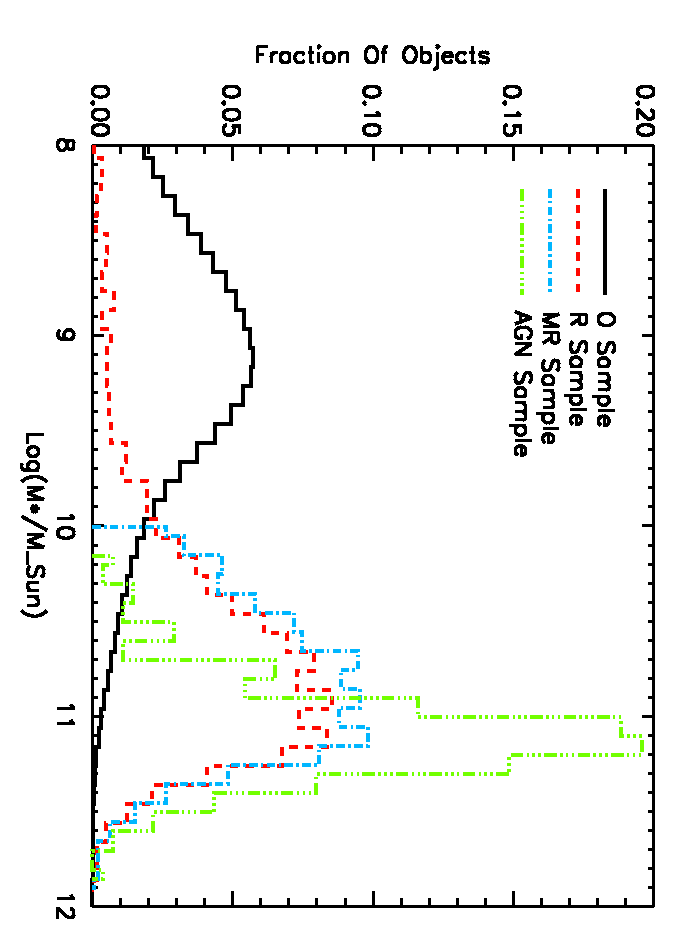}}
\caption{\textit{Stellar Mass Distributions}. The black solid line refers to the O sample, the red dashed one to the R sample, the cyan dot-dashed line refers to the MR sample and the green triple dot-dashed one refers to the AGN sample.}
\label{massdist}
\end{figure}

The greatest difference in the samples is highlighted in Figure \ref{massdist}, which shows the stellar mass distributions for the same samples as in the previous plot. It can be easily seen that the galaxies showing radio emission (R, MR and AGN samples) are found almost only in high mass galaxies (the peak is around $\log(M^{\ast}) \sim 11$), while objects from the O sample span a much more wide range of masses, with a peak at $\log(M^{\ast}) \sim 9$.

This could be due both to an evolutionary effect, that sees the onset of radio emission preferentally in high mass galaxies, or to the limiting flux of the VLA-COSMOS which could limit the kind of galaxies detected to only the most massive and luminous ones. Aside from the physical process responsible for the shape of the stellar mass distributions, the fact that the radio galaxy samples are located at higher masses with respect to normal galaxies must be kept into account, in order not to introduce a bias in the conclusions drawn using these samples. For this reason a cut in stellar mass has been introduced, selecting only the most massive radio sources. Considering only radio sources at $\log(M^{\ast}) \ge 10$ the sample size becomes of 1448 objects, (\textquotedblleft MR sample\textquotedblright). Again, the data analysis has been conducted only on a selection of 1218 with $z \le 2.0$. The cut to $\log(M^{\ast}) \ge 10$ also accounts for the fact that the radio sources without optical identification could 
be faint galaxies with low stellar mass. Therefore selecting only high mass radio sources should avoid the insurgence of any bias in the results from missing optical identifications of radio sources.

\section{The AGN sample}
\label{agndefinition}
It is well known that radio emission could be due, in principle, both to Star Formation (SF) and to AGN phenomena. Several methods have been suggested for dividing the two populations where each of the two phenomena is dominant. There are several recipes for segregating the radio population into sources with emission from an AGN or from star formation (see \textit{e.g.} \citeads{2008ApJS..177...14S, 2009A&A...495..431B, 2005MNRAS.362....9B}, and references therein). In this work two ways which share a common hypothesis have been adopted.

\begin{figure}
\resizebox{\hsize}{!}{\includegraphics[angle=90]{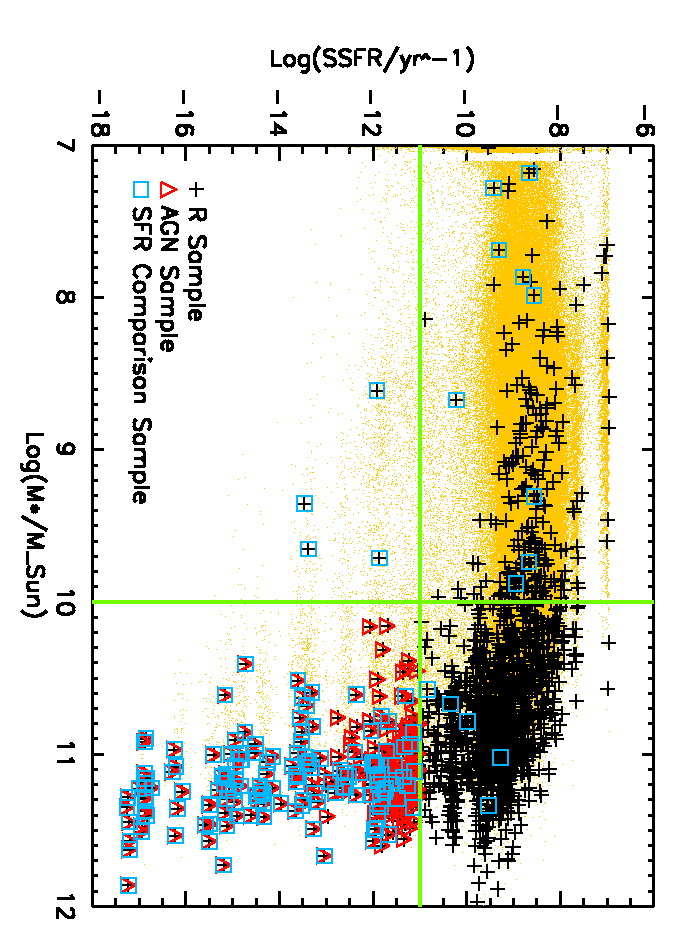}}
\caption{\textit{SSFR - $M^{\ast}$ plane}. The black crosses refer to the R sample, the red triangles to the AGN sample defined through the SSFR cut and the cyan squares to the AGN  sample defined through SFR comparison. The green lines correspond to the cuts in Stellar Mass and Specific Star Formation Rate described in text. In yellow, the O sample is reported for comparison.}
\label{ssfrmstar}
\end{figure}

The first way of defining a sample of radio AGNs is through a cut in Specific Star Formation Rate (SSFR). For each object the SSFR is defined as the ratio of the SFR and the stellar mass, with both quantities derived from the SED fitting procedure. Quiescent galaxies are selected and it is assumed that for this kind of sources all the radio flux comes from AGN activity. For this purpose, the Specific Star Formation Rate \textit{vs} Stellar Mass (SSFR - $M^{\ast}$) plane is analysed (Figure \ref{ssfrmstar}). From this plot, it is clearly visible the locus of star-forming galaxies as an horizontal region within a narrow range of SSFR ($-11 \le \log(SSFR/yr^{-1}) \le -7$). On the right of the figure, a vertical band located at  $10 \le \log(M^{\ast}) \le 12$ represents the galaxies which are switching off  their star formation (\textquotedblleft the dead line\textquotedblright). As AGN sample it has been chosen to select radio sources with $\log(M^{\ast}) \ge 10$ and  $\log(SSFR/yr^{-1}) \le -11$. 
With these cuts, a sample of 276 radio Active Galactic Nuclei has been defined (272 at $z \le 2$).

Although efficient, this method adopts an extreme hypothesis, since it assumes that all radio flux is coming from radio AGN activity in quiescent galaxies. The cuts applied to the sample may actually result inaccurate as more reliable methods to discriminate between radio activity from AGN phenomena and from star formation can be developed.

One of these is described in \citetads{2010A&A...511A...1B}. It consists in the comparison between Star Formation Rates as resulted from the SED fitting of \citetads{2010ApJ...709..644I} and as estimated from the radio power as done in \citetads{2003ApJ...586..794B}. The hypothesis is that all radio flux exceeding that given by the $SFR_{SED}$ is due to AGN activity.

\begin{figure}
\resizebox{\hsize}{!}{\includegraphics[angle=90]{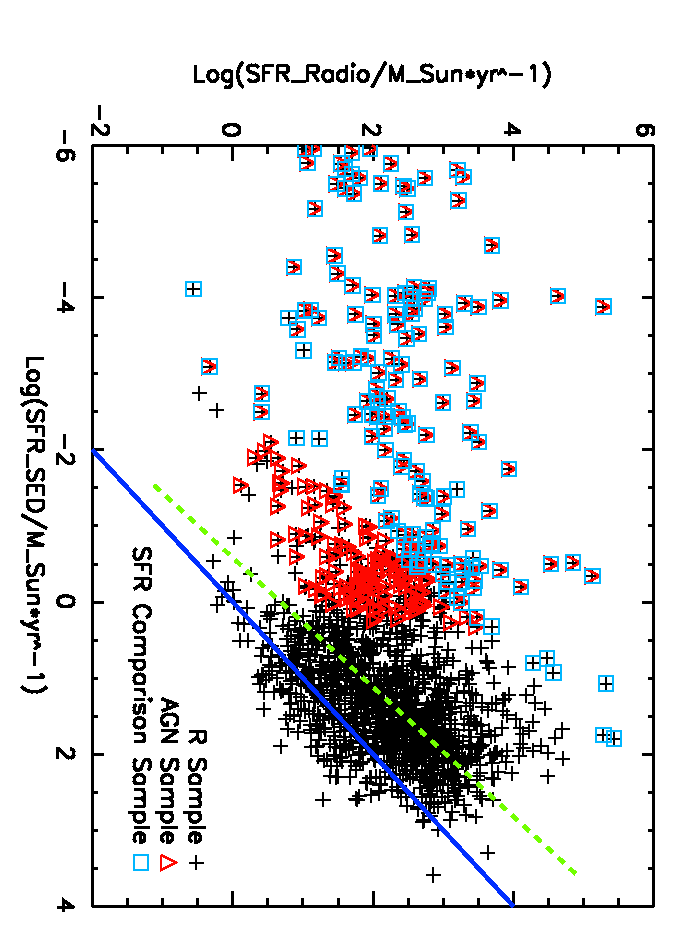}}
\caption{\textit{AGN sample definition}. The black crosses represent the R 
sample, the red triangles represent the AGNs selected through the cut in 
Specific Star Formation Rate, while the cyan squares represent the AGN sample defined through SFR comparison. The dashed green line represents the linear fit relation described in text, while the solid blue line shows the $SFR_{radio} = SFR_{SED}$ relation.}
\label{sfrradiosedclean}
\end{figure}

The $SFR_{Radio}$ has been computed using the \citetads{2003ApJ...586..794B} formula (which is derived for a Salpeter IMF), with a correction of $\sim 2$ in the normalization introduced to be consistent with Herschel-PEP data (Bardelli et al. in prep.). This normalization correction also takes into account the difference between IMFs used in the SFR derivation from optical SED fitting and in the formula from \citetads{2003ApJ...586..794B}. Note that in this particular work the normalization is less important, since only the identification of the locus of star forming galaxies matters. The comparison between the two star formation rates is reported in Figure \ref{sfrradiosedclean}. In this figure it is clearly visible an elongated region due to the star-forming galaxies with a slight zero point shift and tilt. In order to fit this relation, only those objects that have $\log(SFR_{radio}) \le \log(SFR_{SED}) + 2$ have been considered. The linear fit resulted in $\log(SFR_{radio}) = 1.176 \times \log(SFR_{SED}) 
+ 0.685$. The distribution of the distances from the best-fit line has then been fitted with a Gaussian distribution, obtaining a dispersion of $1.40$. In order to be very conservative, all the objects that lie at a distance greater than $3\sigma$ from the fit line are defined as AGNs. With this method, 164 sources are selected (154 at $z \le 2$).

It has to be noted that this second method is more conservative, since the assumptions are less extreme: it is now possible for every galaxy, even for the quiescent ones, that part of the radio emission could be due to residual star formation. The discrimination is set to decide which of the two phenomena (radio AGN activity or the formation of stars) prevails on the other. With this method it is increased the probability of correctly selecting a radio AGN source (purity), in spite of completeness, since some radio AGN sources may not be individuated by the selection criteria. Figures \ref{ssfrmstar} and \ref{sfrradiosedclean} both show that the radio AGN samples derived with the two methods overlap.

Since the infrared emission is a better tracer for the SFR, a more rigorous approach would be to apply the method described above using the SFR derived from IR data, rather than from optical ones. For this reason the R sample has been also matched to the DR1 catalogue of the Herschel PACS Evolutionary Probe\footnote{\url{http://www.mpe.mpg.de/ir/Research/PEP/DR1}} (Herschel-PEP). A description of the survey, observational strategies and data reduction techniques may be found in \citetads{2011A&A...532A..90L}, while SFR and physical quantities have been obtained through a SED fitting procedure performed with a modified version of the \textit{MAGPHYS} code \citepads[see][and references therein for a detailed description]{2013A&A...551A.100B, 2014MNRAS.439.2736D}. A total of 923 sources have been found matching the R sample, 45 of which are in the AGN sample and are all situated at lower SFR values compared with other Herschel sources.

From the comparison of the $SFR_{Radio}$ with the SFR from the IR data, derived through a procedure of SED fitting that allows for the correction of the AGN contribution to galaxy emission ($SFR_{IR}$), a sample of 83 AGNs is created. The method used is equal to the one applied in previous paragraphs: the $SFR_{Radio}$ and $SFR_{IR}$ are compared and the bulk of the relation is linearly fitted in order to isolate sources with an excess of $SFR_{Radio}$ with respect to $SFR_{IR}$. The distribution of the distances of each object from the best-fit line is fitted with a Gaussian distribution and only sources residing farther from the best-fit line than $3\sigma$ are defined as AGNs.

The environmental analysis has been performed on these sources in the same way as for the other AGN samples. The environmental segregation effect is recovered, but unfortunately the sample is too small to grant significativity to the results.

Extensive tests have been performed to check whether the various methods of defining AGNs yield differences in the results found in this work. No significant discrepancies emerge from the tests. The AGN sample defined through the cuts in SSFR and stellar mass will be used since it permits a slightly larger statistics, but all the conclusions that will be derived in the following remain valid also if the radio AGNs are selected through the $SFR_{Radio}$ and $SFR_{SED}$ (or $SFR_{IR}$) comparison.

\section{Method}
\label{method}
The aim of this work is to investigate the presence and properties of galaxy overdensities in the COSMOS Survey field, especially at high redshift. First of all, a definition of overdensity is needed, since many different ways of defining this quantity can be found in the literature \citepads[see]{2013ApJS..206....3S, 2010ApJ...708..505K, 2010A&A...520A..42C} and new ways are still undergoing developement \citepads[see \textit{e.g.}]{2014ApJ...792..113C}. In the present work it has been chosen to have the most simple approach by counting objects in a parallelepipedon with a base side of 1 Mpc (comoving), centred on the considered source, and height twice $\varDelta z = 3 \times \sigma_{\varDelta z/(1+z)} \times (1+z_p)$. It is useful to define three kinds of samples, to which it will be referred in the following.

\begin{description}
\item\textbf{\textit{Starting} sample}. This kind of samples is composed of the objects around which the environment is to be estimated and that will be placed at the centre of each parallelepipedon. \textit{Starting} samples will be the R, MR and AGN samples.

\item\textbf{\textit{Target} sample}. These objects are those that are counted inside each parallelepipedon and that are used to estimate the environment around every galaxy of each \textit{Starting} sample. The O sample will be used as \textit{Target} sample.

\item\textbf{\textit{Control} sample}. The absolute values and richness distributions of overdensities do not have any direct meaning if not compared with those obtained with a {\it Control} sample. In order to create it, for each galaxy of a {\it Starting} sample the redshift has been maintained, but coordinates have been randomatically extracted. The way in which this is achieved is different for every sample and it will be thoroughly explained in section \ref{control}. The environment has then been re-estimated for every \textit{Control} sample using the same {\it Target} sample as for the corresponding real ones. The richness distributions between every sample and its corresponding \textit{Control} one have been compared with a Kolmogorov-Smirnov test. In order not to be influenced by only one extraction this process has been repeated 100 times.
\end{description}

The environmental analysis has been performed in three different redshift bins, namely $z \in [0.0-0.7[$, $[0.7-1.0[$ and $[1.0-2.0]$. It is important to mention that all the environmental analysis has been performed both on a 1 Mpc and on a 2 Mpc scale. It has been found that all the results exposed in the following sections have their significance greatly increased in the 1 Mpc compared to the 2 Mpc case. This could be due to the fact that when adopting a smaller scale only the central regions of the bound structures are considered, especially at high redshfit, where clusters and groups were in an earlier stage of formation. It is known that in such regions (\textit{i.e.} at the centre of overdensities) galaxy formation happens on shorter timescales. Therefore any existing relation between processes involved in galaxy evolution and environment (such as the presence of radio AGNs) is likely to be better evidenced on a 1 Mpc scale. Larger scales (such as the 2 Mpc case) may have the signal diluted by the 
inclusion of regions in structures that are not yet well formed and by an increase in the number of interlopers due to the photometric 
redshfit error. This is confirmed by the fact that the results of this work are much more significant when using the 1 Mpc scale compared to the 2 Mpc case. For this reason, in the following, only the results obtained with an estimate of the environment on a scale of 1 Mpc will be discussed.

As already mentioned in previous sections, several tests have been performed to check whether the choice of the photometric error and of the magnitude cut performed on the sample do affect the results of this work. In particular, the environmental analysis has been performed by using also a cut on the O sample at brighter magnitudes, namely $i^+ < 25.5$, $i^+ < 25$ and $i^+ < 24$. Only the magnitude cut on the O sample has been changed, since no AGN sources have $25 < i^+ < 26.5$. In this way, while the magnitude distribution for the AGN sample is maintained, that of the \textit{Control} samples and of the sources used to trace the environment has been changed. In the first two magnitude cuts the assumption of the photometric redshift error has also been degraded to $\sigma_{\varDelta z/(1+z)} = 0.1 \div 0.2$ in the third redshift bin ($z \in [1.0-2.0]$). The significativity of the results is maintained for all the combinations, except that for the AGNs in the highest redshift bin when using a magnitude 
cut for the O sample of $i^+ < 24$. This is not unexpected, though, since tracing the environment using only the most luminous sources and comparing AGNs with luminous control galaxies has the effect of comparing objects residing (especially at high redshift) in similar environments, therefore reducing the significativity of the signal of environmental segregation. It is therefore possible to conclude that neither the magnitude cut used, nor the value of the photometric redshift error assumed have a systematic effect on the results of this work.
 
\subsection{\textit{Control} samples definition}
\label{control}
As briefly explained above, the \textit{Control} samples for the R, MR and AGN subsamples have been created with optical data. For every galaxy in the R, MR and AGN samples, coordinates have been randomly extracted from the O sample in the same redshift bin for every \textit{Starting} sample. In addition, the following prescriptions have been used separately for every subsample.

\begin{description}
\item\textbf{R Sample}. The \textit{Control} sample RO has been extracted from O, so to have the same number of sources as the R sample in every redshift bin.

\item\textbf{MR Sample}. The \textit{Control} sample MO has been extracted from O, so to have the same number of sources as the MR sample in every redshift bin. Moreover, every galaxy of the \textit{Control} sample has been extracted among those having stellar mass in an interval of $\pm 0.5$ dex from the mass of each galaxy composing the MR sample. In this way, the MO sample has been extracted having the same stellar mass distribution as the MR one.

\item\textbf{AGN Sample}. The \textit{Control} sample QO has been extracted from O, so to have the same number of sources as the AGN sample in every redshift bin. It has also been extracted having the same stellar mass distribution as the AGN sample, with the same method used to extract the MO \textit{Control} sample. Also, the galaxies used as a pool for the QO sample extraction have been selected to be quiescent ($\log(SSFR/yr^{-1}) \le -11$), therefore the QO sample is located in the same lower right region of Figure \ref{ssfrmstar} as the AGN sample.
\end{description}

\section{Results}
\label{results}
In this section the results of the analysis of the environment performed on the various \textit{Starting} samples are analysed. As explained before, in order to better evidentiate the differences in the environment around radio sources and around galaxies with no radio emission (represented by the extracted \textit{Control} samples), the overdensity richness distributions have been compared through the use of a Kolmogorov-Smirnov test, repeated with 100 independent extractions of the various \textit{Control} samples. When applied, the KS test gives the probability that two distributions come from the same parent population. In order to quantify the difference in the environments recovered, only richness distributions that have a KS probability value lower than 0.05 of being extracted from the same parent population will be considered as significative. In Table \ref{ksmezzohundred} the number of \textit{Control} sample extractions (out of 100) that are below the significativity threshold of 0.05 are reported, 
together with the median values of the KS probability value distributions (in brackets).

\begin{table*}
\caption{Number of extractions of the various \textit{Control} samples that are below the significativity threshold of 0.05. In brackets, the median value of the KS probability value distributions is given. The probability resulting from the KS test is the probability of the two distributions analysed being extracted from the same parent population.}
\centering
\begin{tabular}{c c c c}
\hline\hline
Sample    & $0\le z < 0.7$        & $0.7 \le z < 1$        & $1 \le z \le 2$       \\
\hline
R \textit{vs} RO   & 100 ($6.6 \times 10^{-8}$) & 100 ($7.5 \times 10^{-9}$) & 100 ($6.5 \times 10^{-7}$)  \\
MR \textit{vs} MO  & 100 ($6.5 \times 10^{-7}$)  & 100 ($6.9 \times 10^{-7}$) & 100 ($9.2 \times 10^{-6}$)                \\
AGN \textit{vs} QO & 100 ($8.6 \times 10^{-5}$)  & 100 ($6.0 \times 10^{-6}$) & 87 (0.006)  \\
\hline
\end{tabular}
\label{ksmezzohundred}
\end{table*}

\subsection{R Sample}
\label{mezzor}
\begin{figure}
\resizebox{\hsize}{!}{\includegraphics[angle=90]{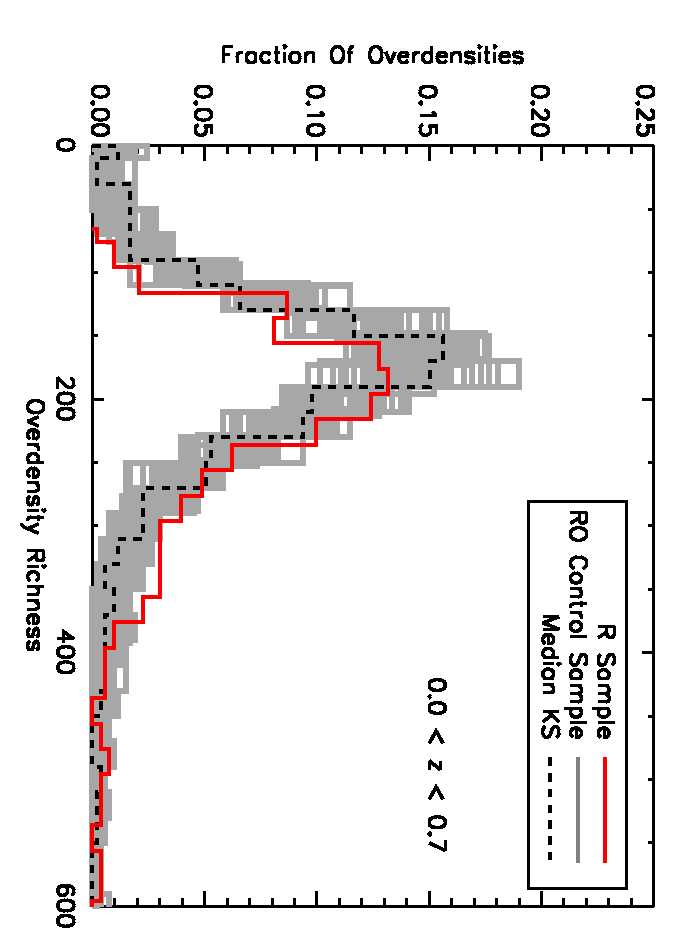}}
\resizebox{\hsize}{!}{\includegraphics[angle=90]{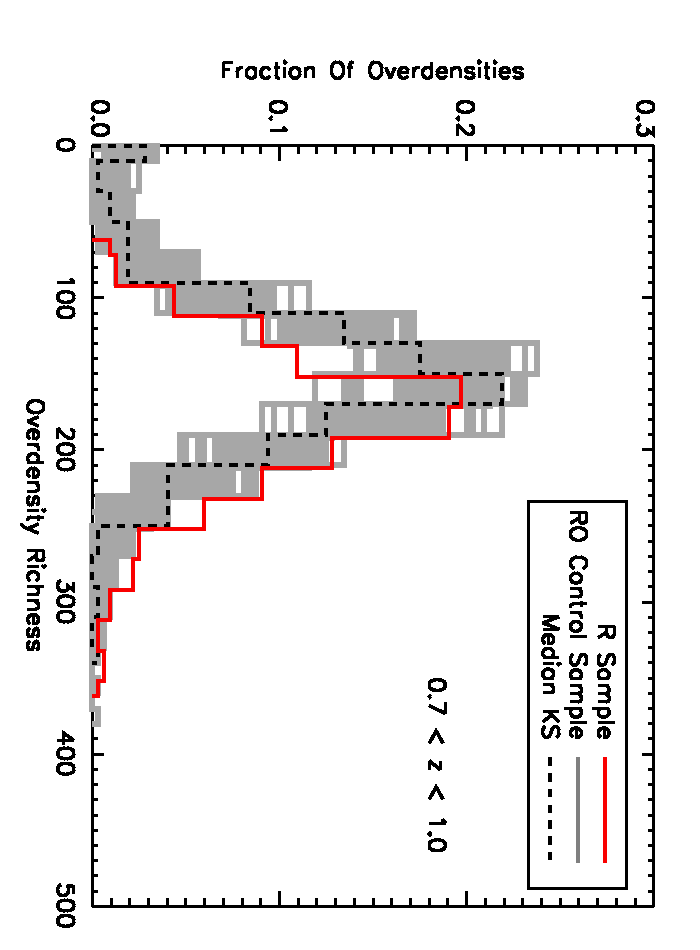}}
\resizebox{\hsize}{!}{\includegraphics[angle=90]{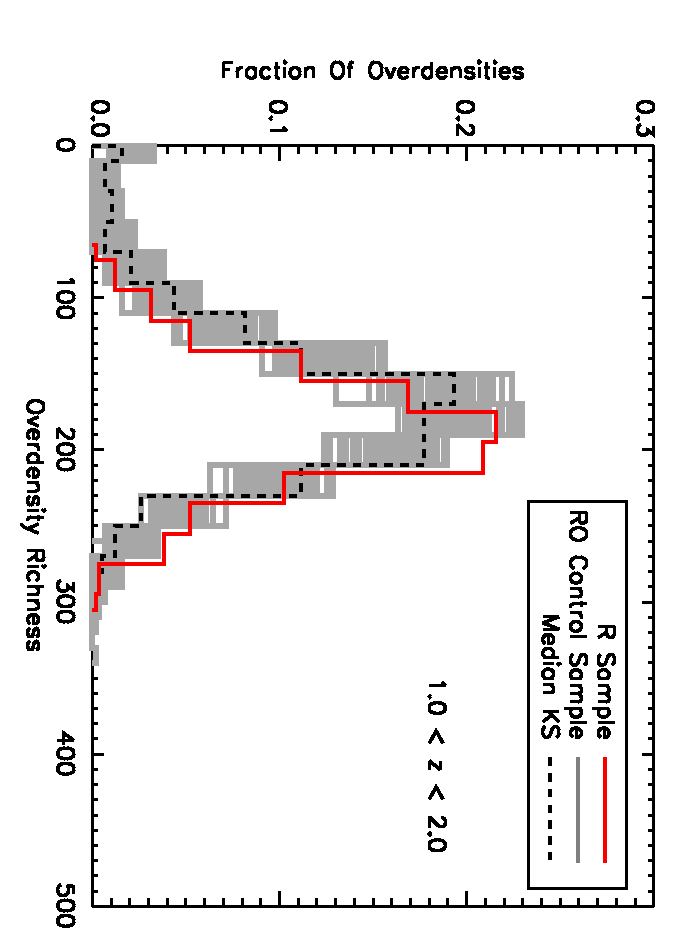}}
\caption{\textit{Galaxy overdensity richness distributions, R and RO sample}. The top panel 
refers to $0 \le z < 0.7$, middle panel to $0.7 \le z < 1$ and the bottom panel to $1 \le z \le 2$. The red solid line refers to the R sample, the grey lines correspond to the 100 extractions of the RO \textit{Control} sample. The dashed black line is the richness distribution of the \textit{Control} sample extraction corresponding to the median value of the KS probability value distribution.}
\label{mezzoerre}
\end{figure}

At first, the environment is estimated around sources from \textit{Starting} samples R and RO. It has to be kept in mind that the R sample is the whole radio catalogue. This sample does not discriminate whether the radio emission comes from star formation or AGN activity. The {\it Target} sample used is represented by the whole O sample. Results of the KS test between R and the 100 RO extractions are listed in the first line of Table \ref{ksmezzohundred}. From these values it can be seen, already, that the environment around radio sources is significantly different from the environment around sources with no sign of radio emission, in every redshift bin. The total of the \textit{Control} sample extractions is always below the significativity threshold, with very low median values. The way in which these environments differ is exemplified by Figure \ref{mezzoerre}. 

In this figure, the overdensity richness distributions for the R sample and for all the 100 extractions of the RO \textit{Control} sample are shown. The richness distribution of the \textit{Control} sample extraction that corresponds to the median of the KS probability value distribution is in evidence. It can be seen that in every redshift bin the radio sources are sistematically distributed at higher overdensity richness than normal galaxies.

Although important, this result could be due in principle to other effects, such as the different mass distributions or evolutionary status of R and RO galaxies. A deeper investigation of the phenomenon is thus needed.

\subsection{M Sample}
\label{massample}
It has been shown (see Figure \ref{massdist}) that optical galaxies and radio sources have different mass distributions. In particular, radio sources are preferentially found in high mass galaxies, which in turn are known to reside preferentally in high density environments. This fact could be the actual reason behind the difference in environmental density found between sources from the R and RO samples, described in the previous section.

In order to rule out this effect, the sample MR and its corresponding \textit{Control} sample MO have been created. These samples are designed to include only massive galaxies ($\log(M^{\ast}) \ge 10$) and to have the same stellar mass distribution. In this way, being the MR sample composed of radio sources and its corresponding \textit{Control} sample MO of optical galaxies, any difference in the environment between the two can be ascribed to the actual presence of radio emission.

\begin{figure}
\resizebox{\hsize}{!}{\includegraphics[angle=90]{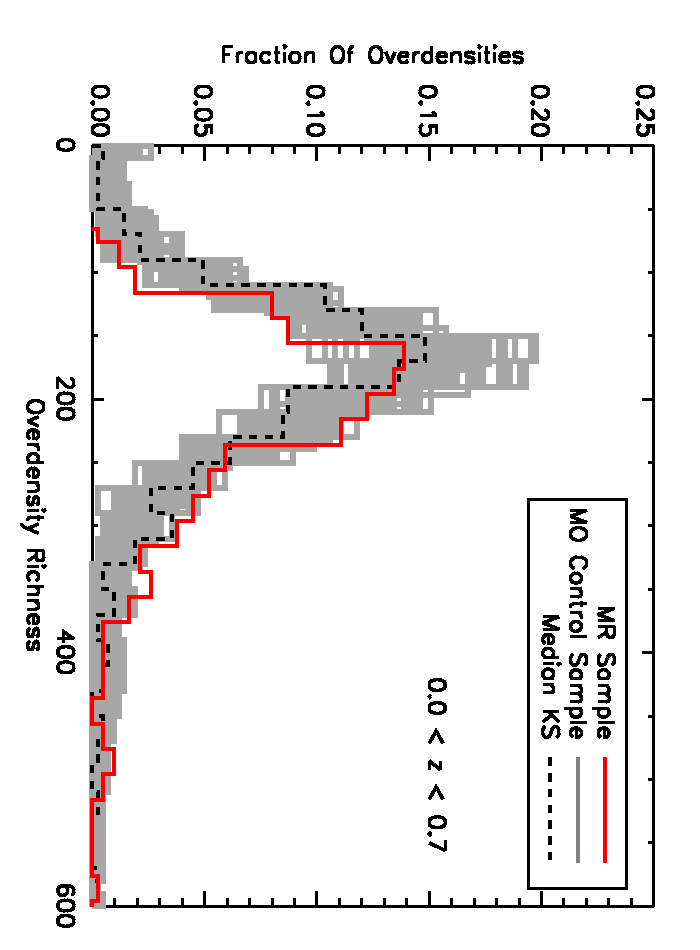}}
\resizebox{\hsize}{!}{\includegraphics[angle=90]{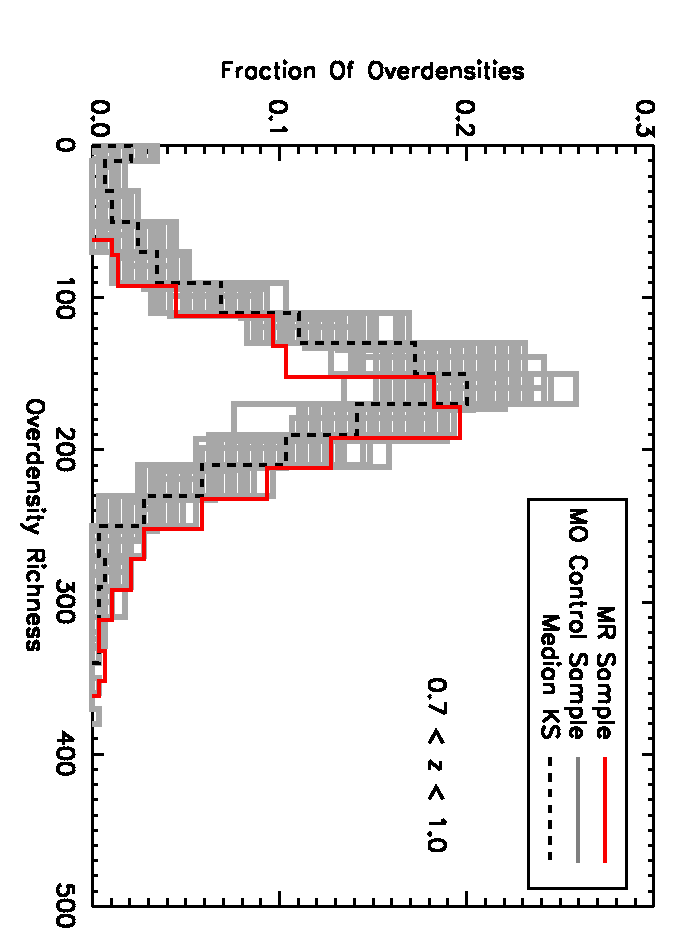}}
\resizebox{\hsize}{!}{\includegraphics[angle=90]{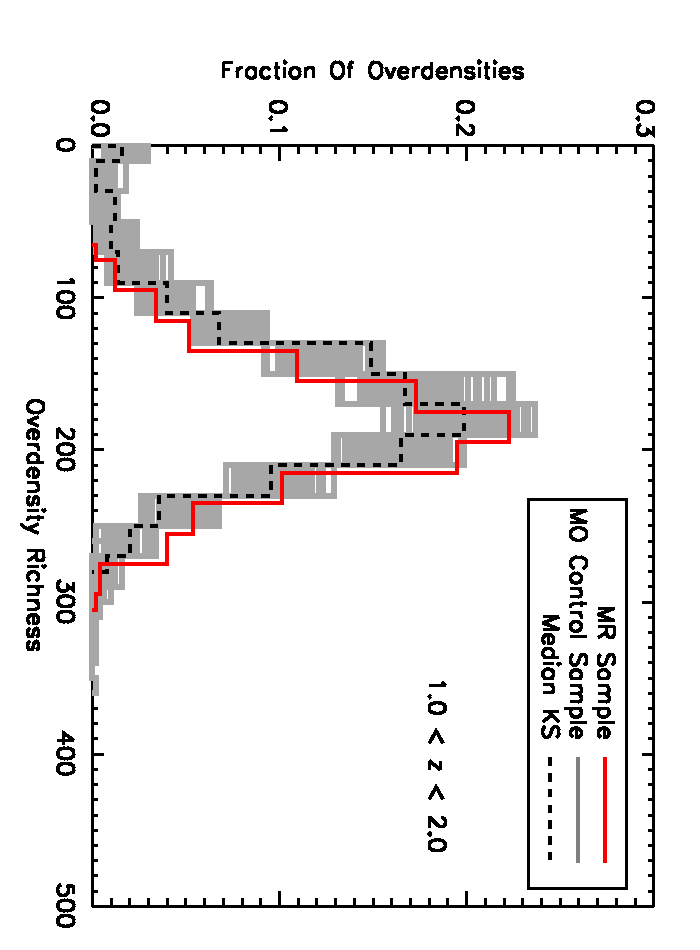}}
\caption{\textit{Galaxy overdensity richness distributions, samples MR and MO}. The top panel refers to $0 \le z < 0.7$, middle panel to $0.7 \le z < 1$ and the bottom panel to $1 \le z \le 2$. The red solid line refers to the MR sample, the grey lines correspond to the 100 extractions of the MO \textit{Control} sample. The dashed black line is the richness distribution of the \textit{Control} sample extraction corresponding to the median value of the KS probability value distribution.}
\label{mezzomass}
\end{figure}

The second line of Table \ref{ksmezzohundred} confirms the trend already found for the R and RO sample. The environment around radio sources and the environment around normal galaxies are significantly different, as demonstrated by the fact that all 100 extractions of the MO \textit{Control} sample are below the significativity threshold, with very low median values. Again a visual inspection of the overdensity richness distributions, as those reported in Figure \ref{mezzomass} for the MR sample and all of the 100 extractions of the MO \textit{Control} sample, confirms that overdensities found around massive radio sources are distributed at higher richness values than those around massive galaxies without radio emission.

It is therefore possible to exclude a stellar mass effect behind the differences in the environment around normal galaxies and radio-emitting ones. Instead these differences are to be ascribed to the physical process behind the radio emission and its correlation with environmental density.

\section{The environment of radio AGNs}
\label{agnenvironment}
The main focus of this paper is the study of the environment around radio AGN sources in the VLA-COSMOS Survey field. This task has been performed in a similar way also in other works \citepads[for example]{2000MNRAS.317..720B,2014MNRAS.445..280H}. In this work a general approach is adopted and the results are in good agreement with the literature.

The radio AGN sample for environment definition has been extracted from the whole population of radio sources in the way described in section \ref{agndefinition}. By using a cut in Specific Star Formation Rate, only sources whose radio emission is due to AGN activity have been selected and their environment has been determined using the O sample as {\it Target} sample.

Nevertheless, the method used to isolate radio AGNs has the effect of selecting also radio sources hosted by massive and quiescent galaxies. These two kind of objects are both naturally found in high density environments, therefore leading to a possible bias in the conclusions drawn. In fact, any difference in the environment around radio AGNs compared to optical galaxies could be actually due to quiescent galaxies being found preferentially inside overdensities or to massive galaxies being the more clustered. In order not to be affected by these problems, the QO \textit{Control} sample has been selected carefully to avoid this bias. Galaxies in the QO sample have been extracted in the same SSFR range and with the same stellar mass distribution of the AGN sample. In this way, every difference in the environmental properties can be ascribed to the presence of radio AGN phenomena.

The results of the comparison of the overdensity richness distributions are shown in the third line of Table \ref{ksmezzohundred}. It can be seen that the environment around radio AGN sources is significantly different from the environment around normal galaxies, with the vast majority of the extractions of the QO sample being under the significativity threshold. Also the median values confirm this trend, being always under $5\%$. This is the most important result of this work, showing a real correlation between environment and the presence of radio AGN phenomena.

\begin{figure}
\resizebox{\hsize}{!}{\includegraphics[angle=90]{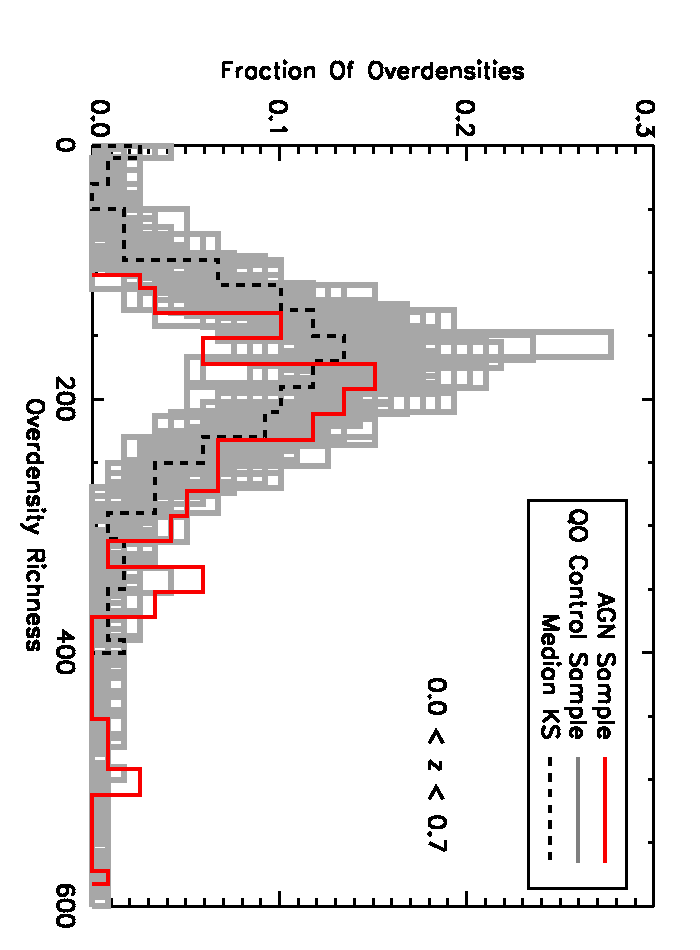}}
\resizebox{\hsize}{!}{\includegraphics[angle=90]{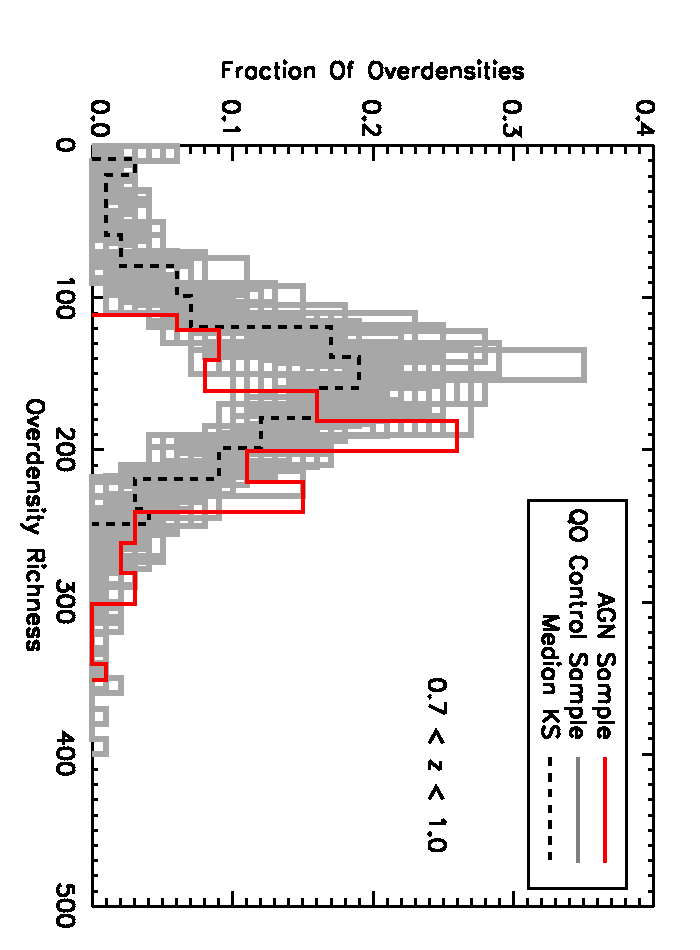}}
\resizebox{\hsize}{!}{\includegraphics[angle=90]{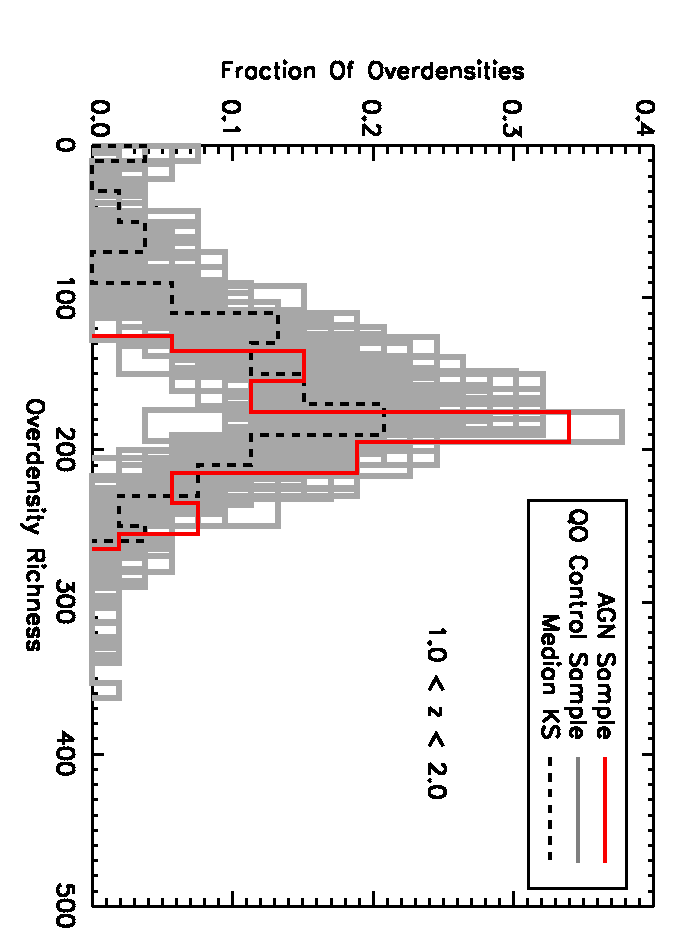}}
\caption{\textit{Galaxy overdensity richness distributions, samples AGN and QO}. The top panel refers to $0 \le z < 0.7$, middle panel to $0.7 \le z < 1$ and the bottom panel to $1 \le z \le 2$. The red solid line refers to the AGN sample, the grey lines correspond to the 100 extractions of the QO \textit{Control} sample. The dashed black line is the richness distribution of the \textit{Control} sample extraction corresponding to the median value of the KS probability value distribution.}
\label{mezzoagn}
\end{figure}

Figure \ref{mezzoagn} shows the overdensity richness distributions for the AGN sample and for the 100 extractions of the QO \textit{Control} sample. It can be clearly seen that overdensities around radio AGNs are distributed at higher richness values than those around normal galaxies. It is therefore possible to conclude that the environment around radio AGNs is significantly richer than around normal galaxies, with this effect being due to radio AGN presence, rather than to radio AGNs being found preferentially in massive or quiescent galaxies (see for example \citeads{2009ApJ...696..891H} and \citeads{1996AJ....112....9L}). In other words, galaxies in denser environments show an enhanced probability of hosting a radio AGN than galaxies in less dense environments.

As a check on the results it has also been estimated the environment around a sample of radio-emitting starforming galaxies, selected in the upper right quadrant of Figure \ref{ssfrmstar}. These sources, characterized by $\log(M^{\ast}) \ge 10$ and  $\log(SSFR/yr^{-1}) \ge -11$, have been compared with a \textit{Control} sample extracted in the same SSFR region and with the same stellar mass distribution. An environmental effect has been found in this sample too, at all redshifts, with star-forming radio sources being in higher density environments. The median values of the KS probability value distribution, though, are higher by an order of magnitude with respect to those in the last line of Table \ref{ksmezzohundred}, except that in the highest redshift bin, even if the sample of star-forming radio sources is $\sim 10$ times greater than the AGN one (450 SF sources against 53 AGNs). Although the great inequality in statistics between radio AGNs and star-forming sources surely plays a major role in 
determining the significativity of the environmental effect for the latter kind of objects, cause of this are also probably residual differences in the Star Formation Rate between the star-forming radio sources and the \textit{Control} sample ones. This is evidenced by the fact that, although they are situated in the same SFR range, the star-forming radio sources are detected by the VLA instrument while the \textit{Control} sample galaxies are not. Therefore the SFR is higher in the radio-detected sample compared to the \textit{Control} one, or there could still be some residual contamination from AGN emission, which induces the environmental effect on these sources.

\subsection{High power and low power radio-sources}
\label{frii}
\begin{figure}
\resizebox{\hsize}{!}{\includegraphics[angle=90]{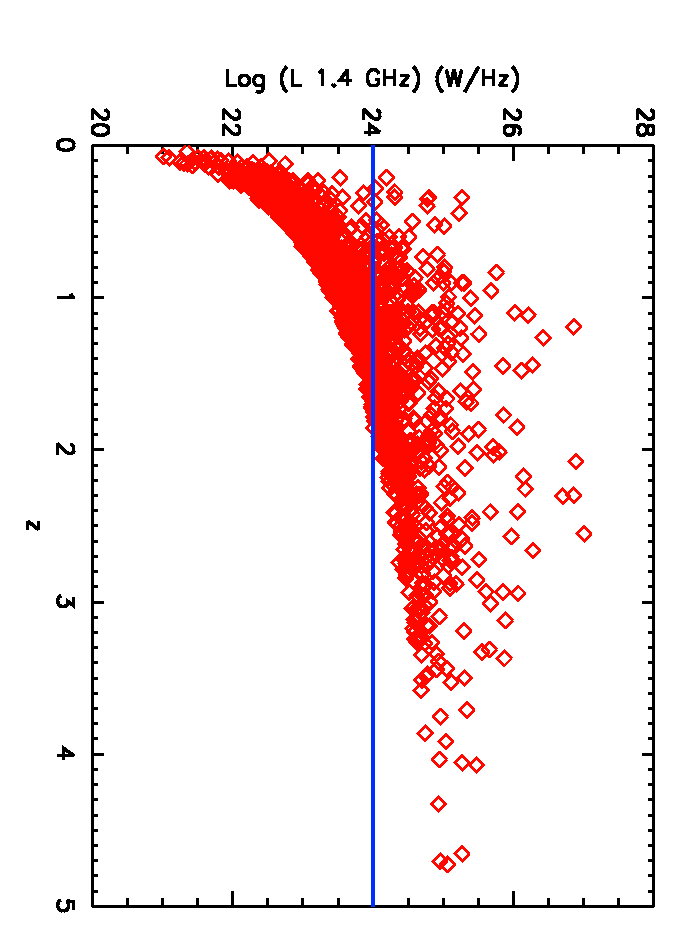}}
\caption{\textit{Redshift dependence of radio luminosity}. Only the R sample is represented.}
\label{malmquist}
\end{figure}

It has been suggested \citepads[see]{2008A&ARv..15...67M, 2010ApJ...710L.107C} that low power radio AGNs are often found in high density regions \citepads[see also]{2014ApJ...792..114C,2009A&A...495..431B,2010MNRAS.407.1078D}. In fact, low power radio AGNs are used to detect protoclusters at high redshifts \citepads[see {\it e.g.}]{2012MNRAS.425..347C, 2013ApJ...769...79W}. In the next it is then explored if the environmental density is actually depending on the radio power. To do this, every sample has been divided in a high power subsample ($\log(L_{1.4 GHz}) \ge 24.5$) and a low power one ($24 \le \log(L_{1.4 GHz}) < 24.5$). The reason for the lower limit at $\log(L_{1.4 GHz}) = 24$ is to avoid the effect induced by the flux limit of the VLA-COSMOS Survey. Such an effect is visible in Figure \ref{malmquist}. 

This figure shows the dependence from redshift of the radio luminosity for the R sample. It can be seen that at high redshifts only the most luminous sources will be detected, due to the survey limiting flux (effect known as the \textit{Malmquist Bias}). This could lead to a bias in the results based on the sample, since more luminous objects could represent a different population of radio sources with respect to less luminous ones. This problem can be solved by considering only the sources more luminous than $\log(L_{1.4 GHz}) = 24$. In this way, a complete sample of radio sources up to $z = 2$ can be created.

It is worth noting that the limit at $\log(L_{1.4 GHz}) = 24.5$, set to divide the high power and low power radio subsamples, roughly corresponds to the canonical division between FRI and FRII objects \citepads{1974MNRAS.167P..31F}.

Therefore the samples AGNH and QOH (where H in the sample name stands for high power) and the samples AGNL and QOL (where L stands for low power) have been created. The number of galaxies in each sample is shown in Table \ref{radionumber}.

\begin{table}
\caption{Number of galaxies in each sample for every redshift bin. High power and low power distinction.}
\centering
\begin{tabular}{c c c c}
\hline\hline
Sample           & $0 \le z < 0.7$        & $0.7 \le z < 1$      & $1 \le z \le 2$   \\
\hline
AGNH and QOH     &    6                   &    16                &    19             \\
AGNL and QOL     &    16                  &    30                &    17             \\
\hline
\end{tabular}
\label{radionumber}
\end{table}

The results of the analysis of the environment in relation with the distinction in radio power are now discussed. In particular, in Table \ref{ksradiohundred} the number of \textit{Control} sample extractions below the significativity level of 0.05 are reported, together with the median values of the KS probability value distribution.

\begin{table}
\caption{Number of extractions of the various \textit{Control} samples that are below the significativity threshold of 0.05. In brackets, the median value of the KS probability value distributions is given. Low power and high power radio source distinction.}
\centering
\begin{tabular}{c c c c}
\hline\hline
Sample          & $0 \le z < 0.7$       & $0.7 \le z < 1$       & $1 \le z \le 2$        \\
\hline
AGNH \textit{vs} QOH     & 9 (0.08)     & 43 (0.07)       & 17 (0.25)                  \\
AGNL \textit{vs} QOL     & 93 (0.002)  & 99 (0.0006)     & 37 (0.08)                   \\
\hline
\end{tabular}
\label{ksradiohundred}
\end{table}

\begin{figure}
\resizebox{\hsize}{!}{\includegraphics[angle=90]{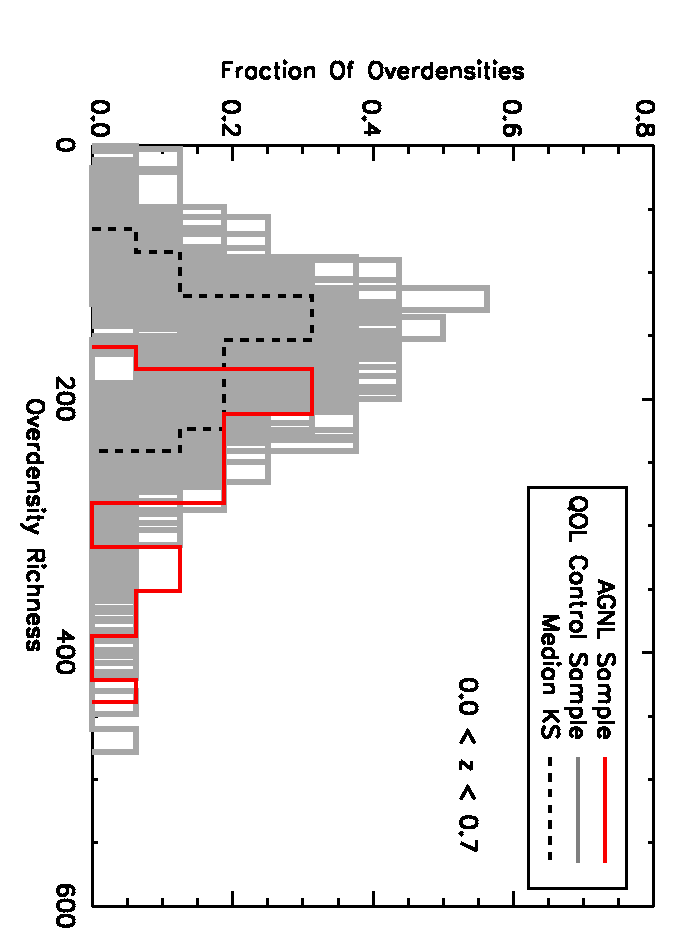}}
\resizebox{\hsize}{!}{\includegraphics[angle=90]{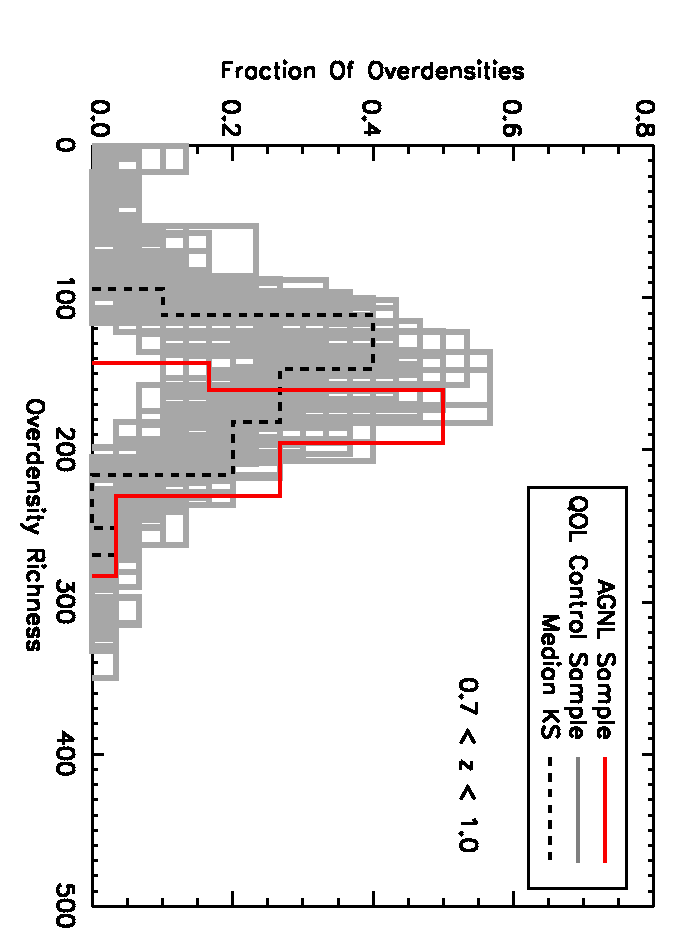}}
\resizebox{\hsize}{!}{\includegraphics[angle=90]{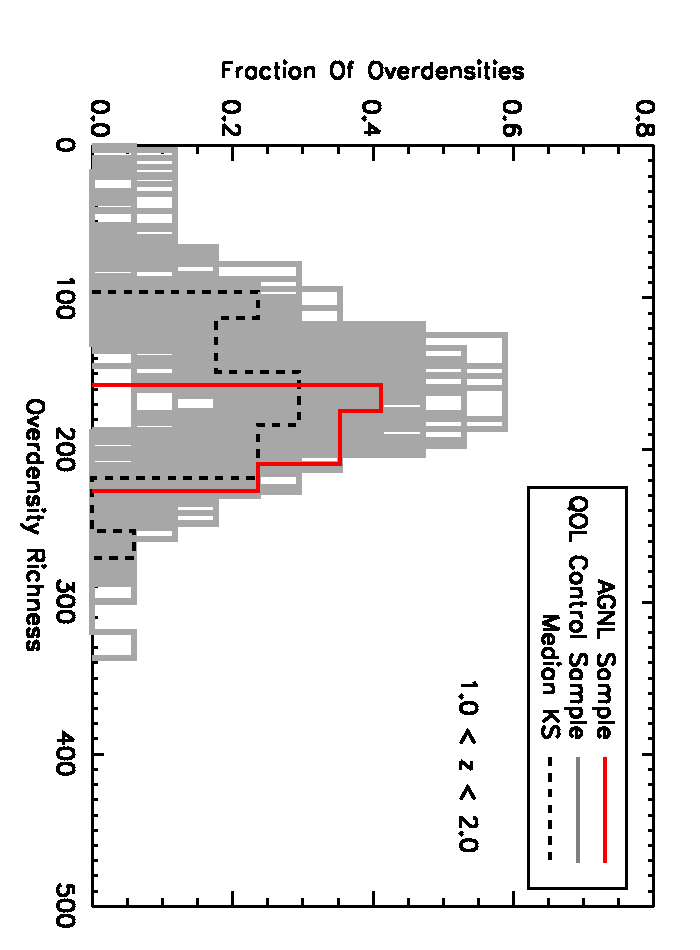}}
\caption{\textit{Galaxy overdensity richness distributions, samples AGNL and QOL}. The top panel refers to $0 \le z < 0.7$, the middle panel to $0.7 \le z < 1$ and the bottom panel to $1 \le z \le 2$. The red solid line refers to the AGNL sample, the grey lines correspond to the 100 extractions of the QOL \textit{Control} sample. The dashed black line is the richness distribution of the \textit{Control} sample extraction corresponding to the median value of the KS probability value distribution.}
\label{mezzoagnlow}
\end{figure}

It is possible to see that the high power and low power samples have different KS probabilities: the environmental effect remains present only for low power radio sources. In fact, in Table \ref{ksradiohundred} it can be seen that the number of extractions below the significativity level is fairly low for the high power radio AGNs, never reaching even $50\%$ of the total, and the median KS probability value is always greater than 0.05. On the other hand, for low power radio AGNs the number of extractions below the significativity threshold is close to $100\%$. Only in the highest redshift bin the value is quite low ($37\%$), but it is nevertheless larger than the corresponding value in the same redshift bin for the high power AGNs. In Figure \ref{mezzoagnlow} overdensity richness distributions are shown for the case of the AGNL sample and for the extractions of the corresponding QOL \textit{Control} sample. The distributions of the AGNL sample being sistematically shifted towards higher richness values 
confirm the trend. 

This is another important result of this work: when considering the environment of radio sources, low power AGNs are found preferentially in high density environments. Therefore, higher overdensity richness values enhance the probability of a galaxy to host a low power radio AGN. This result is not unexpected. In fact, it could arise as a direct consequence of different accretion modes for low power and high power AGNs. In particular, theory predicts low power radio AGNs to be fuelled by hot gas (such as the one found at the center of clusters where cooling flows are present), whereas high power AGNs would be fuelled by cold gas \citepads[see for example][and references therein]{2007MNRAS.376.1849H}. If this were true, then low power AGNs (fuelled by hot gas) would be expected to reside at the centre of rich overdensities, such as clusters and groups (where hot gas is present). On the other hand, high power AGNs would not be found in high density environments, both for the lack of cold gas necessary for 
accretion and for the inefficiency of wet mergers (due to the high velocity dispersions of galaxies at the centre of clusters and rich overdensities), which are the trigger mechanism that has been proposed for the ignition of high power radio AGNs. The results of this work seem to 
support this view of AGN and galaxy formation.

\section{Correlation with known clusters and groups catalogues}
\label{correlationclusters}

In the previous sections it has been found that radio AGNs live in environments which are on average denser than those of normal galaxies. It has therefore been explored whether these overdensities are related to virialized or bound objects. Two compilations of clusters and groups from the literature have been used: that of \citetads{2007ApJS..172..182F}, found in the COSMOS field analysing deep Chandra and XMM observations (see references in the paper), and the zCOSMOS group catalogue \citepads{2012ApJ...753..121K}. For this comparison, only the two most distant redshift bins of the AGN sample have been considered ($ z \in [0.7-1.0[$ and $z \in [1.0-2.0]$). At those redshifts, a 0.5 Mpc value corresponds approximately to 71 ($z = 0.7$), 63 ($z = 1.0$) and 61 ($z = 2.0$) arcsec. However, the largest source of uncertainties is coming from the large error in the photometric redshifts.

To be conservative, all objects that are within a radius of $\sim 60$ arcsec and $|\varDelta z|< 0.12$ for $0.7 \le z < 1.0$ and $\sim 60$ arcsec and $|\varDelta z|< 0.2$ for $1.0 \le z \le 2.0$ from the cluster locations reported in the two catalogues have been considered, in order to take into account the varying error in the photometric redshifts estimate.

In Table \ref{Finoguenov}, the 26 associations with the \citetads{2007ApJS..172..182F} sample are listed, while in Table \ref{knobel} those with the zCOSMOS group catalogue of \citetads{2012ApJ...753..121K} are reported. It has been found that $\sim 10-20 \%$ of the AGN sample is in groups and clusters. Furthermore, no significant difference between the radio power distributions of the radio AGNs in clusters and groups and of the radio sources of this work is detected. The only noticeable source is the one with $\log(L_{1.4 GHz}) = 26.86$ in a cluster in the most distant redshift bin, which is also the brightest object in the AGN sample.

For what concerns the X-ray luminosities, there is a slight tendency of clusters hosting a radio AGN at $z \in [0.7-1.0[$ to be brighter, with a KS probability of being extracted by the same population of $\sim 0.09$. However the statistics is not sufficient to say anything for the highest redshift bin.

In the case of zCOSMOS groups, the richness distributions are significantly different, with groups hosting a radio AGN being richer (KS probability of $\sim 10^{-3}$). This holds also when considering only groups with more than 3 elements.

\section {Mass functions}
\label{massfunctions}
It is now investigated whether there is a relative evolution of the mass functions of radio AGNs and those of normal galaxies. To perform this task, stellar masses as determined in the UltraVISTA Survey \citepads{2012A&A...544A.156M,2013A&A...556A..55I} are used. A limiting magnitude for the UltraVISTA Survey of $K_S = 24$ is assumed in the data. Moreover, the radio power limit of $\log(L_{1.4 GHz}) \ge 24$ previously introduced has also been applied to the AGN sample (leading to a total of 88 sources), since lack of completeness in the radio data could bias the estimate of the redshift evolution of the mass functions.

\begin{figure}
\resizebox{\hsize}{!}{\includegraphics[angle=90]{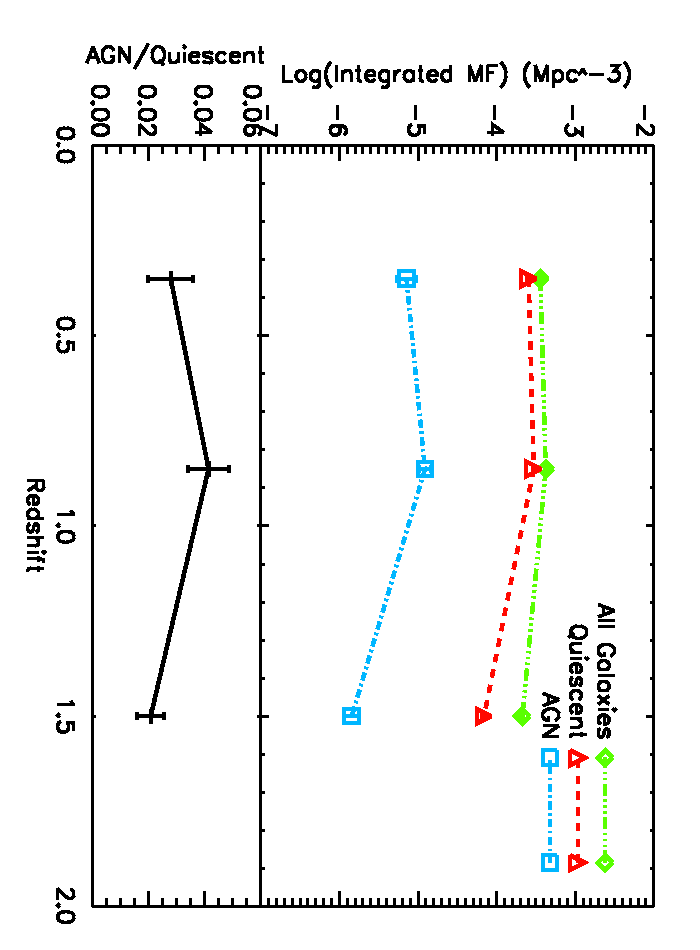}}
\caption{\textit{Redshift evolution of the integrated mass functions}. Upper panel: value of the integrated mass functions. The curves refer to all UltraVISTA galaxies (green diamonds), a selection of only the quiescent ones (red triangles) and the radio AGN sample (cyan squares). Lower panel: ratio of the integrated mass function values for AGN hosting and quiescent galaxies. Points have been placed at values of redshift corresponding to the mean redshift of each redshift bin.}
\label{IMFredshiftEVO}
\end{figure}

Stellar mass functions are computed with the non-parametric $1/V_{max}$ estimator (\citeads{1980ApJ...235..694A}, see also \citeads{2002A&A...395..443B} and references therein for further information). For radio objects, the $V_{max}$ is calculated taking into account the different radio and $K_S$ limiting magnitudes of the surveys. In particular the minimum between the $V_{max}$ set by radio and optical data has been used. Mass functions have been calculated for all UltraVISTA galaxies, a subsample of quiescent ones (selected through a cut to $\log(SSFR/yr^{-1}) \le -11$) and the AGN sample. The size of the AGN sample is too small to draw conclusions as a function of mass, therefore only the integral of the stellar mass functions for galaxies more massive than $\log(M^{\ast}) = 11$ is considered (this new cut further reduces the sample to 71 sources). These results are shown in Figure \ref{IMFredshiftEVO}. Error bars are computed by simply summing all the $1/V^2_{max}$ terms due to Poisson statistics, as 
explained in \citetads{2002A&A...395..443B}.

Although the integrated mass function for all galaxies varies, within a factor $\sim 2$, between the redshift bins $[1.0-2.0]$ and $[0.7-1.0[$ the corresponding variation for the integrated mass function of quiescent and radio AGNs is even stronger (a factor $\sim 5$ and almost an order of magnitude for the two samples respectively). In these same redshift bins the ratio of radio AGN hosting to quiescent galaxies increases from $\sim 0.02$ to $\sim 0.04$. This increase could confirm that the presence of a radio AGN is a phenomenon that evolves with quiescent galaxies, but at a different rate.

In order to explore if the ratio of radio AGNs to quiescent galaxies evolves also as a function of the local density, all samples have been divided in two, considering objects above and below the median of the overdensity richness distributions of all galaxies as obtained with our method. With these cuts, the numbers of objects in the AGN sample in the high/low density regions are, respectively, 13/3, 33/11,17/10 for the low, middle and high redshift bins. For the quiescent sample, the statistics is, instead, always in the 2000-3000 range in every redshift bin.

\begin{figure}
\resizebox{\hsize}{!}{\includegraphics[angle=90]{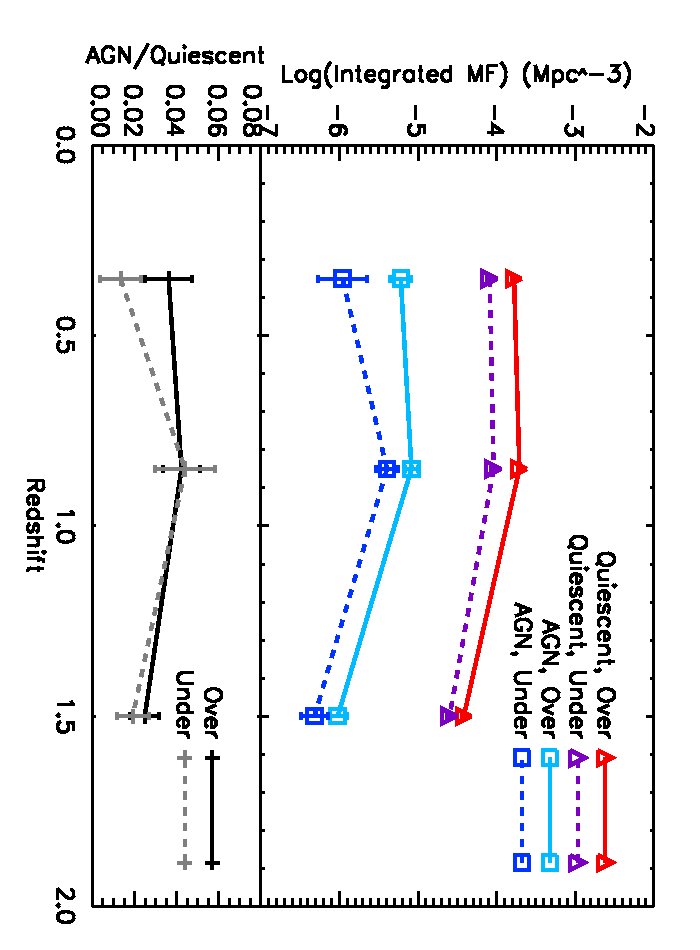}}
\caption{\textit{Environment effect on the integrated mass functions}. All samples have been split in two, as stated by the legend, with the terms \textquotedblleft Over\textquotedblright$\:$and \textquotedblleft Under\textquotedblright$\:$referring to high density environments (over the median value) and low density ones (under the median value). Upper panel: red and purple triangles refer to quiescent UltraVISTA galaxies, cyan and blue squares to the AGN sample. Lower panel: black and grey curves represent the ratio of AGN hosting to quiescent galaxies. Solid lines are for high density environments, dashed lines for the low density ones. Points have been placed at values of redshift corresponding to the mean redshift of each redshift bin.}
\label{IMFenvironmentEVO}
\end{figure}

As expected, the objects from the AGN sample are sistematically more numerous above the median value than below. This could be an indication of the effect previously discovered in a more significant way, of radio AGNs being found in higher density environments. The effect of the environment on the integrated mass functions and on the ratio of AGN hosting to quiescent galaxies is showed in Figure \ref{IMFenvironmentEVO}.

The ratio of AGN hosting galaxies to quiescent ones shows that the environmental segregation is present. In particular the ratio is larger in high density environments compared to low density ones at $z \le 0.7$. In the first redshift bin there is a factor $\sim 4$ difference between high and low density environments. This seems to indicate that the environmental segregation is somewhat more important for radio AGNs. Also, while the evolution is similar, both the integrated mass function of the AGN sample and the ratio of AGN to quiescent galaxies show an enhanced probability of a galaxy to host a radio AGN in high density environments in the lowest redshift bin. This result is in very good agreement with what has been found in previous sections with other methods.

\section{Conclusions and summary}
\label{conclusions}
The aim of this work was to investigate the role of the environment in determining the presence and the properties of radio AGN phenomena. These phenomena are predicted and needed by the theory of galaxy formation, therefore understanding the way in which environment influences their upcoming could lead to a better comprehension of the processes through which galaxies are formed. The deep photometric redshift sample extracted from the COSMOS survey \citepads{2009ApJ...690.1236I} has been used, together with the VLA-COSMOS survey \citepads{2007ApJS..172...46S} for the radio data.

From these surveys, various samples have been extracted, with particular focus on AGN samples. Results may be summarized as follows:

\begin{enumerate}
\item
Analysing the R sample and the RO \textit{Control} sample, it is found that the environment where radio sources reside is significantly denser than the environment around galaxies without radio emission.

\item
These results are not due to radio sources being hosted by high mass galaxies (which in turn are known to reside in denser environments). In fact, by repeating the environment estimate on samples designed to have the same stellar mass distributions (MR and MO samples) the effect of radio sources residing in sistematically and significantly denser environments still holds.

\item
A sample of radio AGNs has been extracted from the whole catalogue of radio sources. It is found that the environment around radio AGNs is significantly denser than the environment around normal galaxies in the same stellar mass and Specific Star Formation Rate range (QO sample). This shows that the environmental segregation that is found is actually due to an enhanced probability of galaxies to host a radio AGN in denser environments. This leads to conclude that environment plays a role in determining the onset of AGN phenomena in galaxies, and that these pehnomena are important in the process of galaxy formation.

\item
The environmental effect in enhancing the probability of a galaxy to host a radio AGN is dominated by low power radio AGNs. In fact, the difference in the environment between low power radio AGNs and normal galaxies is significant, while for high power radio AGNs it is not. This has been found by splitting all the previously analysed samples according to a cut in their radio luminosity. This effect, of low power radio AGNs showing the only signal of environemntal segregation, could be related to the thermodynamic properties of the gas fuelled to the AGN \citepads[see for example]{2007MNRAS.376.1849H}.

\item
Analysing the integrated mass functions, calculated for the whole population of the UltraVISTA galaxies and sub-samples of only quiescent and radio AGN hosting ones, hints of a redshift evolution can be found. This can be seen both in the decrease of the value of the integrated mass functions with redshift for the samples of quiescent galaxies and radio AGNs, as well as in the same decrease with redshift in their ratio.

\item
Exploring the dependence from the environment of the integrated mass functions, it is found that the fraction of radio AGNs is always enhanced in high density environments, in agreement with the stronger results found using richness distributions. Moreover, radio AGNs are also the population with the greatest difference in the values of integrated mass functions between high and low density environments.
\end{enumerate}

The picture that emerges from this work is that high density environment is indeed related to the presence and the properties of radio AGNs. This fact is in good agreement with the current theory of galaxy formation, that predicts galaxies in high density environments to undergo star formation quenching faster and in a more efficient way. Future studies and the increase in the statistics for the mass functions could open the way to a more thorough analysis of the existing correlations.

\begin{acknowledgements}
Part of this work is based on data products from observations made with ESO Telescopes at the La Silla Paranal Observatory under ESO programme ID 179.A-2005 and on data products produced by TERAPIX and the Cambridge Astronomy Survey Unit on behalf of the UltraVISTA consortium. We thank the anonymous referee for the helpful comments. Reproduced with permission from Astronomy \& Astrophysics, \textcopyright ESO.
\end{acknowledgements}

\bibliographystyle{aa}
\bibliography{Malavasi}

\begin{thebibliography}{52}
\expandafter\ifx\csname natexlab\endcsname\relax\def\natexlab#1{#1}\fi

\bibitem[{{Avni} \& {Bahcall}(1980)}]{1980ApJ...235..694A}
{Avni}, Y. \& {Bahcall}, J.~N. 1980, \apj, 235, 694

\bibitem[{{Bardelli} {et~al.}(2010){Bardelli}, {Schinnerer}, {Smol{\v c}ic},
  {Zamorani}, {Zucca}, {Mignoli}, {Halliday}, {Kova{\v c}}, {Ciliegi},
  {Caputi}, {Koekemoer}, {Bongiorno}, {Bondi}, {Bolzonella}, {Vergani},
  {Pozzetti}, {Carollo}, {Contini}, {Kneib}, {Le F{\`e}vre}, {Lilly},
  {Mainieri}, {Renzini}, {Scodeggio}, {Coppa}, {Cucciati}, {de la Torre}, {de
  Ravel}, {Franzetti}, {Garilli}, {Iovino}, {Kampczyk}, {Knobel}, {Lamareille},
  {Le Borgne}, {Le Brun}, {Maier}, {Pell{\`o}}, {Peng}, {Perez-Montero},
  {Ricciardelli}, {Silverman}, {Tanaka}, {Tasca}, {Tresse}, {Abbas}, {Bottini},
  {Cappi}, {Cassata}, {Cimatti}, {Guzzo}, {Leauthaud}, {Maccagni}, {Marinoni},
  {McCracken}, {Memeo}, {Meneux}, {Oesch}, {Porciani}, {Scaramella}, {Capak},
  {Sanders}, {Scoville}, {Taniguchi}, \& {Jahnke}}]{2010A&A...511A...1B}
{Bardelli}, S., {Schinnerer}, E., {Smol{\v c}ic}, V., {et~al.} 2010, \aap, 511,
  A1

\bibitem[{{Bardelli} {et~al.}(2009){Bardelli}, {Zucca}, {Bolzonella},
  {Ciliegi}, {Gregorini}, {Zamorani}, {Bondi}, {Zanichelli}, {Tresse},
  {Vergani}, {Gavignaud}, {Bongiorno}, {Bottini}, {Garilli}, {Le Brun}, {Le
  F{\`e}vre}, {Maccagni}, {Scaramella}, {Scodeggio}, {Vettolani}, {Adami},
  {Arnouts}, {Cappi}, {Charlot}, {Contini}, {Foucaud}, {Franzetti}, {Guzzo},
  {Ilbert}, {Iovino}, {Lamareille}, {McCracken}, {Marano}, {Marinoni},
  {Mazure}, {Meneux}, {Merighi}, {Paltani}, {Pell{\`o}}, {Pollo}, {Pozzetti},
  {Radovich}, {Abbas}, {Brinchmann}, {Cucciati}, {de La Torre}, {de Ravel},
  {Memeo}, {Perez-Montero}, {Mellier}, {Merluzzi}, {Temporin}, {de Ruiter}, \&
  {Parma}}]{2009A&A...495..431B}
{Bardelli}, S., {Zucca}, E., {Bolzonella}, M., {et~al.} 2009, \aap, 495, 431

\bibitem[{{Bell}(2003)}]{2003ApJ...586..794B}
{Bell}, E.~F. 2003, \apj, 586, 794

\bibitem[{{Berta} {et~al.}(2013){Berta}, {Lutz}, {Santini}, {Wuyts}, {Rosario},
  {Brisbin}, {Cooray}, {Franceschini}, {Gruppioni}, {Hatziminaoglou}, {Hwang},
  {Le Floc'h}, {Magnelli}, {Nordon}, {Oliver}, {Page}, {Popesso}, {Pozzetti},
  {Pozzi}, {Riguccini}, {Rodighiero}, {Roseboom}, {Scott}, {Symeonidis},
  {Valtchanov}, {Viero}, \& {Wang}}]{2013A&A...551A.100B}
{Berta}, S., {Lutz}, D., {Santini}, P., {et~al.} 2013, \aap, 551, A100

\bibitem[{{Best}(2000)}]{2000MNRAS.317..720B}
{Best}, P.~N. 2000, \mnras, 317, 720

\bibitem[{{Best}(2004)}]{2004MNRAS.351...70B}
{Best}, P.~N. 2004, \mnras, 351, 70

\bibitem[{{Best} {et~al.}(2005){Best}, {Kauffmann}, {Heckman}, \&
  {Ivezi{\'c}}}]{2005MNRAS.362....9B}
{Best}, P.~N., {Kauffmann}, G., {Heckman}, T.~M., \& {Ivezi{\'c}}, {\v Z}.
  2005, \mnras, 362, 9

\bibitem[{{Bolzonella} {et~al.}(2010){Bolzonella}, {Kova{\v c}}, {Pozzetti},
  {Zucca}, {Cucciati}, {Lilly}, {Peng}, {Iovino}, {Zamorani}, {Vergani},
  {Tasca}, {Lamareille}, {Oesch}, {Caputi}, {Kampczyk}, {Bardelli}, {Maier},
  {Abbas}, {Knobel}, {Scodeggio}, {Carollo}, {Contini}, {Kneib}, {Le
  F{\`e}vre}, {Mainieri}, {Renzini}, {Bongiorno}, {Coppa}, {de la Torre}, {de
  Ravel}, {Franzetti}, {Garilli}, {Le Borgne}, {Le Brun}, {Mignoli},
  {Pell{\'o}}, {Perez-Montero}, {Ricciardelli}, {Silverman}, {Tanaka},
  {Tresse}, {Bottini}, {Cappi}, {Cassata}, {Cimatti}, {Guzzo}, {Koekemoer},
  {Leauthaud}, {Maccagni}, {Marinoni}, {McCracken}, {Memeo}, {Meneux},
  {Porciani}, {Scaramella}, {Aussel}, {Capak}, {Halliday}, {Ilbert},
  {Kartaltepe}, {Salvato}, {Sanders}, {Scarlata}, {Scoville}, {Taniguchi}, \&
  {Thompson}}]{2010A&A...524A..76B}
{Bolzonella}, M., {Kova{\v c}}, K., {Pozzetti}, L., {et~al.} 2010, \aap, 524,
  A76

\bibitem[{{Bolzonella} {et~al.}(2002){Bolzonella}, {Pell{\'o}}, \&
  {Maccagni}}]{2002A&A...395..443B}
{Bolzonella}, M., {Pell{\'o}}, R., \& {Maccagni}, D. 2002, \aap, 395, 443

\bibitem[{{Brusa} {et~al.}(2007){Brusa}, {Zamorani}, {Comastri}, {Hasinger},
  {Cappelluti}, {Civano}, {Finoguenov}, {Mainieri}, {Salvato}, {Vignali},
  {Elvis}, {Fiore}, {Gilli}, {Impey}, {Lilly}, {Mignoli}, {Silverman}, {Trump},
  {Urry}, {Bender}, {Capak}, {Huchra}, {Kneib}, {Koekemoer}, {Leauthaud},
  {Lehmann}, {Massey}, {Matute}, {McCarthy}, {McCracken}, {Rhodes}, {Scoville},
  {Taniguchi}, \& {Thompson}}]{2007ApJS..172..353B}
{Brusa}, M., {Zamorani}, G., {Comastri}, A., {et~al.} 2007, \apjs, 172, 353

\bibitem[{{Capak} {et~al.}(2007){Capak}, {Aussel}, {Ajiki}, {McCracken},
  {Mobasher}, {Scoville}, {Shopbell}, {Taniguchi}, {Thompson}, {Tribiano},
  {Sasaki}, {Blain}, {Brusa}, {Carilli}, {Comastri}, {Carollo}, {Cassata},
  {Colbert}, {Ellis}, {Elvis}, {Giavalisco}, {Green}, {Guzzo}, {Hasinger},
  {Ilbert}, {Impey}, {Jahnke}, {Kartaltepe}, {Kneib}, {Koda}, {Koekemoer},
  {Komiyama}, {Leauthaud}, {Le Fevre}, {Lilly}, {Liu}, {Massey}, {Miyazaki},
  {Murayama}, {Nagao}, {Peacock}, {Pickles}, {Porciani}, {Renzini}, {Rhodes},
  {Rich}, {Salvato}, {Sanders}, {Scarlata}, {Schiminovich}, {Schinnerer},
  {Scodeggio}, {Sheth}, {Shioya}, {Tasca}, {Taylor}, {Yan}, \&
  {Zamorani}}]{2007ApJS..172...99C}
{Capak}, P., {Aussel}, H., {Ajiki}, M., {et~al.} 2007, \apjs, 172, 99

\bibitem[{{Castignani} {et~al.}(2014{\natexlab{a}}){Castignani}, {Chiaberge},
  {Celotti}, \& {Norman}}]{2014ApJ...792..113C}
{Castignani}, G., {Chiaberge}, M., {Celotti}, A., \& {Norman}, C.
  2014{\natexlab{a}}, \apj, 792, 113

\bibitem[{{Castignani} {et~al.}(2014{\natexlab{b}}){Castignani}, {Chiaberge},
  {Celotti}, {Norman}, \& {De Zotti}}]{2014ApJ...792..114C}
{Castignani}, G., {Chiaberge}, M., {Celotti}, A., {Norman}, C., \& {De Zotti},
  G. 2014{\natexlab{b}}, \apj, 792, 114

\bibitem[{{Chabrier}(2003)}]{2003PASP..115..763C}
{Chabrier}, G. 2003, \pasp, 115, 763

\bibitem[{{Chiaberge} {et~al.}(2010){Chiaberge}, {Capetti}, {Macchetto},
  {Rosati}, {Tozzi}, \& {Tremblay}}]{2010ApJ...710L.107C}
{Chiaberge}, M., {Capetti}, A., {Macchetto}, F.~D., {et~al.} 2010, \apjl, 710,
  L107

\bibitem[{{Chuter} {et~al.}(2011){Chuter}, {Almaini}, {Hartley}, {McLure},
  {Dunlop}, {Foucaud}, {Conselice}, {Simpson}, {Cirasuolo}, \&
  {Bradshaw}}]{2011MNRAS.413.1678C}
{Chuter}, R.~W., {Almaini}, O., {Hartley}, W.~G., {et~al.} 2011, \mnras, 413,
  1678

\bibitem[{{Ciliegi} {et~al.}(2005){Ciliegi}, {Zamorani}, {Bondi}, {Pozzetti},
  {Bolzonella}, {Gregorini}, {Garilli}, {Iovino}, {McCracken}, {Mellier},
  {Radovich}, {de Ruiter}, {Parma}, {Bottini}, {Le Brun}, {Le F{\`e}vre},
  {Maccagni}, {Picat}, {Scaramella}, {Scodeggio}, {Tresse}, {Vettolani},
  {Zanichelli}, {Adami}, {Arnaboldi}, {Arnouts}, {Bardelli}, {Cappi},
  {Charlot}, {Contini}, {Foucaud}, {Franzetti}, {Guzzo}, {Ilbert}, {Marano},
  {Marinoni}, {Mathez}, {Mazure}, {Meneux}, {Merighi}, {Merluzzi}, {Paltani},
  {Pollo}, {Zucca}, {Bongiorno}, {Busarello}, {Gavignaud}, {Pell{\`o}},
  {Ripepi}, \& {Rizzo}}]{2005A&A...441..879C}
{Ciliegi}, P., {Zamorani}, G., {Bondi}, M., {et~al.} 2005, \aap, 441, 879

\bibitem[{{Cooke} {et~al.}(2012){Cooke}, {Pettini}, \&
  {Murphy}}]{2012MNRAS.425..347C}
{Cooke}, R., {Pettini}, M., \& {Murphy}, M.~T. 2012, \mnras, 425, 347

\bibitem[{{Croton} {et~al.}(2006){Croton}, {Springel}, {White}, {De Lucia},
  {Frenk}, {Gao}, {Jenkins}, {Kauffmann}, {Navarro}, \&
  {Yoshida}}]{2006MNRAS.365...11C}
{Croton}, D.~J., {Springel}, V., {White}, S.~D.~M., {et~al.} 2006, \mnras, 365,
  11

\bibitem[{{Cucciati} {et~al.}(2010{\natexlab{a}}){Cucciati}, {Iovino}, {Kova{\v
  c}}, {Scodeggio}, {Lilly}, {Bolzonella}, {Bardelli}, {Vergani}, {Tasca},
  {Zucca}, {Zamorani}, {Pozzetti}, {Knobel}, {Oesch}, {Lamareille}, {Caputi},
  {Kampczyk}, {Tresse}, {Maier}, {Carollo}, {Contini}, {Kneib}, {Le F{\`e}vre},
  {Mainieri}, {Renzini}, {Bongiorno}, {Coppa}, {de la Torre}, {de Ravel},
  {Franzetti}, {Garilli}, {Le Borgne}, {Le Brun}, {Mignoli}, {Pell{\`o}},
  {Peng}, {Perez-Montero}, {Ricciardelli}, {Silverman}, {Tanaka}, {Koekemoer},
  {Scoville}, {Abbas}, {Bottini}, {Cappi}, {Cassata}, {Cimatti}, {Guzzo},
  {Leauthaud}, {Maccagni}, {Marinoni}, {McCracken}, {Memeo}, {Meneux},
  {Porciani}, \& {Scaramella}}]{2010A&A...524A...2C}
{Cucciati}, O., {Iovino}, A., {Kova{\v c}}, K., {et~al.} 2010{\natexlab{a}},
  \aap, 524, A2

\bibitem[{{Cucciati} {et~al.}(2010{\natexlab{b}}){Cucciati}, {Marinoni},
  {Iovino}, {Bardelli}, {Adami}, {Mazure}, {Scodeggio}, {Maccagni}, {Temporin},
  {Zucca}, {De Lucia}, {Blaizot}, {Garilli}, {Meneux}, {Zamorani}, {Le
  F{\`e}vre}, {Cappi}, {Guzzo}, {Bottini}, {Le Brun}, {Tresse}, {Vettolani},
  {Zanichelli}, {Arnouts}, {Bolzonella}, {Charlot}, {Ciliegi}, {Contini},
  {Foucaud}, {Franzetti}, {Gavignaud}, {Ilbert}, {Lamareille}, {McCracken},
  {Marano}, {Merighi}, {Paltani}, {Pell{\`o}}, {Pollo}, {Pozzetti}, {Vergani},
  \& {P{\'e}rez-Montero}}]{2010A&A...520A..42C}
{Cucciati}, O., {Marinoni}, C., {Iovino}, A., {et~al.} 2010{\natexlab{b}},
  \aap, 520, A42

\bibitem[{{Delvecchio} {et~al.}(2014){Delvecchio}, {Gruppioni}, {Pozzi},
  {Berta}, {Zamorani}, {Cimatti}, {Lutz}, {Scott}, {Vignali}, {Cresci},
  {Feltre}, {Cooray}, {Vaccari}, {Fritz}, {Le Floc'h}, {Magnelli}, {Popesso},
  {Oliver}, {Bock}, {Carollo}, {Contini}, {Le F{\'e}vre}, {Lilly}, {Mainieri},
  {Renzini}, \& {Scodeggio}}]{2014MNRAS.439.2736D}
{Delvecchio}, I., {Gruppioni}, C., {Pozzi}, F., {et~al.} 2014, \mnras, 439,
  2736

\bibitem[{{Donoso} {et~al.}(2010){Donoso}, {Li}, {Kauffmann}, {Best}, \&
  {Heckman}}]{2010MNRAS.407.1078D}
{Donoso}, E., {Li}, C., {Kauffmann}, G., {Best}, P.~N., \& {Heckman}, T.~M.
  2010, \mnras, 407, 1078

\bibitem[{{Dressler}(1980)}]{1980ApJ...236..351D}
{Dressler}, A. 1980, \apj, 236, 351

\bibitem[{{Fanaroff} \& {Riley}(1974)}]{1974MNRAS.167P..31F}
{Fanaroff}, B.~L. \& {Riley}, J.~M. 1974, \mnras, 167, 31P

\bibitem[{{Finoguenov} {et~al.}(2007){Finoguenov}, {Guzzo}, {Hasinger},
  {Scoville}, {Aussel}, {B{\"o}hringer}, {Brusa}, {Capak}, {Cappelluti},
  {Comastri}, {Giodini}, {Griffiths}, {Impey}, {Koekemoer}, {Kneib},
  {Leauthaud}, {Le F{\`e}vre}, {Lilly}, {Mainieri}, {Massey}, {McCracken},
  {Mobasher}, {Murayama}, {Peacock}, {Sakelliou}, {Schinnerer}, {Silverman},
  {Smol{\v c}i{\'c}}, {Taniguchi}, {Tasca}, {Taylor}, {Trump}, \&
  {Zamorani}}]{2007ApJS..172..182F}
{Finoguenov}, A., {Guzzo}, L., {Hasinger}, G., {et~al.} 2007, \apjs, 172, 182

\bibitem[{{Granato} {et~al.}(2004){Granato}, {De Zotti}, {Silva}, {Bressan}, \&
  {Danese}}]{2004ApJ...600..580G}
{Granato}, G.~L., {De Zotti}, G., {Silva}, L., {Bressan}, A., \& {Danese}, L.
  2004, \apj, 600, 580

\bibitem[{{Hardcastle} {et~al.}(2007){Hardcastle}, {Evans}, \&
  {Croston}}]{2007MNRAS.376.1849H}
{Hardcastle}, M.~J., {Evans}, D.~A., \& {Croston}, J.~H. 2007, \mnras, 376,
  1849

\bibitem[{{Hatch} {et~al.}(2014){Hatch}, {Wylezalek}, {Kurk}, {Stern}, {De
  Breuck}, {Jarvis}, {Galametz}, {Gonzalez}, {Hartley}, {Mortlock}, {Seymour},
  \& {Stevens}}]{2014MNRAS.445..280H}
{Hatch}, N.~A., {Wylezalek}, D., {Kurk}, J.~D., {et~al.} 2014, \mnras, 445, 280

\bibitem[{{Hickox} {et~al.}(2009){Hickox}, {Jones}, {Forman}, {Murray},
  {Kochanek}, {Eisenstein}, {Jannuzi}, {Dey}, {Brown}, {Stern}, {Eisenhardt},
  {Gorjian}, {Brodwin}, {Narayan}, {Cool}, {Kenter}, {Caldwell}, \&
  {Anderson}}]{2009ApJ...696..891H}
{Hickox}, R.~C., {Jones}, C., {Forman}, W.~R., {et~al.} 2009, \apj, 696, 891

\bibitem[{{Ilbert} {et~al.}(2009){Ilbert}, {Capak}, {Salvato}, {Aussel},
  {McCracken}, {Sanders}, {Scoville}, {Kartaltepe}, {Arnouts}, {Le Floc'h},
  {Mobasher}, {Taniguchi}, {Lamareille}, {Leauthaud}, {Sasaki}, {Thompson},
  {Zamojski}, {Zamorani}, {Bardelli}, {Bolzonella}, {Bongiorno}, {Brusa},
  {Caputi}, {Carollo}, {Contini}, {Cook}, {Coppa}, {Cucciati}, {de la Torre},
  {de Ravel}, {Franzetti}, {Garilli}, {Hasinger}, {Iovino}, {Kampczyk},
  {Kneib}, {Knobel}, {Kovac}, {Le Borgne}, {Le Brun}, {F{\`e}vre}, {Lilly},
  {Looper}, {Maier}, {Mainieri}, {Mellier}, {Mignoli}, {Murayama}, {Pell{\`o}},
  {Peng}, {P{\'e}rez-Montero}, {Renzini}, {Ricciardelli}, {Schiminovich},
  {Scodeggio}, {Shioya}, {Silverman}, {Surace}, {Tanaka}, {Tasca}, {Tresse},
  {Vergani}, \& {Zucca}}]{2009ApJ...690.1236I}
{Ilbert}, O., {Capak}, P., {Salvato}, M., {et~al.} 2009, \apj, 690, 1236

\bibitem[{{Ilbert} {et~al.}(2013){Ilbert}, {McCracken}, {Le F{\`e}vre},
  {Capak}, {Dunlop}, {Karim}, {Renzini}, {Caputi}, {Boissier}, {Arnouts},
  {Aussel}, {Comparat}, {Guo}, {Hudelot}, {Kartaltepe}, {Kneib}, {Krogager},
  {Le Floc'h}, {Lilly}, {Mellier}, {Milvang-Jensen}, {Moutard}, {Onodera},
  {Richard}, {Salvato}, {Sanders}, {Scoville}, {Silverman}, {Taniguchi},
  {Tasca}, {Thomas}, {Toft}, {Tresse}, {Vergani}, {Wolk}, \&
  {Zirm}}]{2013A&A...556A..55I}
{Ilbert}, O., {McCracken}, H.~J., {Le F{\`e}vre}, O., {et~al.} 2013, \aap, 556,
  A55

\bibitem[{{Ilbert} {et~al.}(2010){Ilbert}, {Salvato}, {Le Floc'h}, {Aussel},
  {Capak}, {McCracken}, {Mobasher}, {Kartaltepe}, {Scoville}, {Sanders},
  {Arnouts}, {Bundy}, {Cassata}, {Kneib}, {Koekemoer}, {Le F{\`e}vre}, {Lilly},
  {Surace}, {Taniguchi}, {Tasca}, {Thompson}, {Tresse}, {Zamojski}, {Zamorani},
  \& {Zucca}}]{2010ApJ...709..644I}
{Ilbert}, O., {Salvato}, M., {Le Floc'h}, E., {et~al.} 2010, \apj, 709, 644

\bibitem[{{Knobel} {et~al.}(2012){Knobel}, {Lilly}, {Iovino}, {Kova{\v c}},
  {Bschorr}, {Presotto}, {Oesch}, {Kampczyk}, {Carollo}, {Contini}, {Kneib},
  {Le Fevre}, {Mainieri}, {Renzini}, {Scodeggio}, {Zamorani}, {Bardelli},
  {Bolzonella}, {Bongiorno}, {Caputi}, {Cucciati}, {de la Torre}, {de Ravel},
  {Franzetti}, {Garilli}, {Lamareille}, {Le Borgne}, {Le Brun}, {Maier},
  {Mignoli}, {Pello}, {Peng}, {Perez Montero}, {Silverman}, {Tanaka}, {Tasca},
  {Tresse}, {Vergani}, {Zucca}, {Barnes}, {Bordoloi}, {Cappi}, {Cimatti},
  {Coppa}, {Koekemoer}, {L{\'o}pez-Sanjuan}, {McCracken}, {Moresco}, {Nair},
  {Pozzetti}, \& {Welikala}}]{2012ApJ...753..121K}
{Knobel}, C., {Lilly}, S.~J., {Iovino}, A., {et~al.} 2012, \apj, 753, 121

\bibitem[{{Kova{\v c}} {et~al.}(2010){Kova{\v c}}, {Lilly}, {Cucciati},
  {Porciani}, {Iovino}, {Zamorani}, {Oesch}, {Bolzonella}, {Knobel},
  {Finoguenov}, {Peng}, {Carollo}, {Pozzetti}, {Caputi}, {Silverman}, {Tasca},
  {Scodeggio}, {Vergani}, {Scoville}, {Capak}, {Contini}, {Kneib}, {Le
  F{\`e}vre}, {Mainieri}, {Renzini}, {Bardelli}, {Bongiorno}, {Coppa}, {de la
  Torre}, {de Ravel}, {Franzetti}, {Garilli}, {Guzzo}, {Kampczyk},
  {Lamareille}, {Le Borgne}, {Le Brun}, {Maier}, {Mignoli}, {Pello}, {Perez
  Montero}, {Ricciardelli}, {Tanaka}, {Tresse}, {Zucca}, {Abbas}, {Bottini},
  {Cappi}, {Cassata}, {Cimatti}, {Fumana}, {Koekemoer}, {Maccagni}, {Marinoni},
  {McCracken}, {Memeo}, {Meneux}, \& {Scaramella}}]{2010ApJ...708..505K}
{Kova{\v c}}, K., {Lilly}, S.~J., {Cucciati}, O., {et~al.} 2010, \apj, 708, 505

\bibitem[{{Ledlow} \& {Owen}(1996)}]{1996AJ....112....9L}
{Ledlow}, M.~J. \& {Owen}, F.~N. 1996, \aj, 112, 9

\bibitem[{{Lilly} {et~al.}(2009){Lilly}, {Le Brun}, {Maier}, {Mainieri},
  {Mignoli}, {Scodeggio}, {Zamorani}, {Carollo}, {Contini}, {Kneib}, {Le
  F{\`e}vre}, {Renzini}, {Bardelli}, {Bolzonella}, {Bongiorno}, {Caputi},
  {Coppa}, {Cucciati}, {de la Torre}, {de Ravel}, {Franzetti}, {Garilli},
  {Iovino}, {Kampczyk}, {Kovac}, {Knobel}, {Lamareille}, {Le Borgne}, {Pello},
  {Peng}, {P{\'e}rez-Montero}, {Ricciardelli}, {Silverman}, {Tanaka}, {Tasca},
  {Tresse}, {Vergani}, {Zucca}, {Ilbert}, {Salvato}, {Oesch}, {Abbas},
  {Bottini}, {Capak}, {Cappi}, {Cassata}, {Cimatti}, {Elvis}, {Fumana},
  {Guzzo}, {Hasinger}, {Koekemoer}, {Leauthaud}, {Maccagni}, {Marinoni},
  {McCracken}, {Memeo}, {Meneux}, {Porciani}, {Pozzetti}, {Sanders},
  {Scaramella}, {Scarlata}, {Scoville}, {Shopbell}, \&
  {Taniguchi}}]{2009ApJS..184..218L}
{Lilly}, S.~J., {Le Brun}, V., {Maier}, C., {et~al.} 2009, \apjs, 184, 218

\bibitem[{{Lilly} {et~al.}(2007){Lilly}, {Le F{\`e}vre}, {Renzini}, {Zamorani},
  {Scodeggio}, {Contini}, {Carollo}, {Hasinger}, {Kneib}, {Iovino}, {Le Brun},
  {Maier}, {Mainieri}, {Mignoli}, {Silverman}, {Tasca}, {Bolzonella},
  {Bongiorno}, {Bottini}, {Capak}, {Caputi}, {Cimatti}, {Cucciati}, {Daddi},
  {Feldmann}, {Franzetti}, {Garilli}, {Guzzo}, {Ilbert}, {Kampczyk}, {Kovac},
  {Lamareille}, {Leauthaud}, {Borgne}, {McCracken}, {Marinoni}, {Pello},
  {Ricciardelli}, {Scarlata}, {Vergani}, {Sanders}, {Schinnerer}, {Scoville},
  {Taniguchi}, {Arnouts}, {Aussel}, {Bardelli}, {Brusa}, {Cappi}, {Ciliegi},
  {Finoguenov}, {Foucaud}, {Franceschini}, {Halliday}, {Impey}, {Knobel},
  {Koekemoer}, {Kurk}, {Maccagni}, {Maddox}, {Marano}, {Marconi}, {Meneux},
  {Mobasher}, {Moreau}, {Peacock}, {Porciani}, {Pozzetti}, {Scaramella},
  {Schiminovich}, {Shopbell}, {Smail}, {Thompson}, {Tresse}, {Vettolani},
  {Zanichelli}, \& {Zucca}}]{2007ApJS..172...70L}
{Lilly}, S.~J., {Le F{\`e}vre}, O., {Renzini}, A., {et~al.} 2007, \apjs, 172,
  70

\bibitem[{{Lutz} {et~al.}(2011){Lutz}, {Poglitsch}, {Altieri}, {Andreani},
  {Aussel}, {Berta}, {Bongiovanni}, {Brisbin}, {Cava}, {Cepa}, {Cimatti},
  {Daddi}, {Dominguez-Sanchez}, {Elbaz}, {F{\"o}rster Schreiber}, {Genzel},
  {Grazian}, {Gruppioni}, {Harwit}, {Le Floc'h}, {Magdis}, {Magnelli},
  {Maiolino}, {Nordon}, {P{\'e}rez Garc{\'{\i}}a}, {Popesso}, {Pozzi},
  {Riguccini}, {Rodighiero}, {Saintonge}, {Sanchez Portal}, {Santini}, {Shao},
  {Sturm}, {Tacconi}, {Valtchanov}, {Wetzstein}, \&
  {Wieprecht}}]{2011A&A...532A..90L}
{Lutz}, D., {Poglitsch}, A., {Altieri}, B., {et~al.} 2011, \aap, 532, A90

\bibitem[{{McCracken} {et~al.}(2012){McCracken}, {Milvang-Jensen}, {Dunlop},
  {Franx}, {Fynbo}, {Le F{\`e}vre}, {Holt}, {Caputi}, {Goranova}, {Buitrago},
  {Emerson}, {Freudling}, {Hudelot}, {L{\'o}pez-Sanjuan}, {Magnard}, {Mellier},
  {M{\o}ller}, {Nilsson}, {Sutherland}, {Tasca}, \&
  {Zabl}}]{2012A&A...544A.156M}
{McCracken}, H.~J., {Milvang-Jensen}, B., {Dunlop}, J., {et~al.} 2012, \aap,
  544, A156

\bibitem[{{Miley} \& {De Breuck}(2008)}]{2008A&ARv..15...67M}
{Miley}, G. \& {De Breuck}, C. 2008, \aapr, 15, 67

\bibitem[{{Peng} {et~al.}(2010){Peng}, {Lilly}, {Kova{\v c}}, {Bolzonella},
  {Pozzetti}, {Renzini}, {Zamorani}, {Ilbert}, {Knobel}, {Iovino}, {Maier},
  {Cucciati}, {Tasca}, {Carollo}, {Silverman}, {Kampczyk}, {de Ravel},
  {Sanders}, {Scoville}, {Contini}, {Mainieri}, {Scodeggio}, {Kneib}, {Le
  F{\`e}vre}, {Bardelli}, {Bongiorno}, {Caputi}, {Coppa}, {de la Torre},
  {Franzetti}, {Garilli}, {Lamareille}, {Le Borgne}, {Le Brun}, {Mignoli},
  {Perez Montero}, {Pello}, {Ricciardelli}, {Tanaka}, {Tresse}, {Vergani},
  {Welikala}, {Zucca}, {Oesch}, {Abbas}, {Barnes}, {Bordoloi}, {Bottini},
  {Cappi}, {Cassata}, {Cimatti}, {Fumana}, {Hasinger}, {Koekemoer},
  {Leauthaud}, {Maccagni}, {Marinoni}, {McCracken}, {Memeo}, {Meneux}, {Nair},
  {Porciani}, {Presotto}, \& {Scaramella}}]{2010ApJ...721..193P}
{Peng}, Y.-j., {Lilly}, S.~J., {Kova{\v c}}, K., {et~al.} 2010, \apj, 721, 193

\bibitem[{{Quadri} {et~al.}(2012){Quadri}, {Williams}, {Franx}, \&
  {Hildebrandt}}]{2012ApJ...744...88Q}
{Quadri}, R.~F., {Williams}, R.~J., {Franx}, M., \& {Hildebrandt}, H. 2012,
  \apj, 744, 88

\bibitem[{{Schinnerer} {et~al.}(2010){Schinnerer}, {Sargent}, {Bondi}, {Smol{\v
  c}i{\'c}}, {Datta}, {Carilli}, {Bertoldi}, {Blain}, {Ciliegi}, {Koekemoer},
  \& {Scoville}}]{2010ApJS..188..384S}
{Schinnerer}, E., {Sargent}, M.~T., {Bondi}, M., {et~al.} 2010, \apjs, 188, 384

\bibitem[{{Schinnerer} {et~al.}(2007){Schinnerer}, {Smol{\v c}i{\'c}},
  {Carilli}, {Bondi}, {Ciliegi}, {Jahnke}, {Scoville}, {Aussel}, {Bertoldi},
  {Blain}, {Impey}, {Koekemoer}, {Le Fevre}, \& {Urry}}]{2007ApJS..172...46S}
{Schinnerer}, E., {Smol{\v c}i{\'c}}, V., {Carilli}, C.~L., {et~al.} 2007,
  \apjs, 172, 46

\bibitem[{{Scoville} {et~al.}(2013){Scoville}, {Arnouts}, {Aussel}, {Benson},
  {Bongiorno}, {Bundy}, {Calvo}, {Capak}, {Carollo}, {Civano}, {Dunlop},
  {Elvis}, {Faisst}, {Finoguenov}, {Fu}, {Giavalisco}, {Guo}, {Ilbert},
  {Iovino}, {Kajisawa}, {Kartaltepe}, {Leauthaud}, {Le F{\`e}vre}, {LeFloch},
  {Lilly}, {Liu}, {Manohar}, {Massey}, {Masters}, {McCracken}, {Mobasher},
  {Peng}, {Renzini}, {Rhodes}, {Salvato}, {Sanders}, {Sarvestani}, {Scarlata},
  {Schinnerer}, {Sheth}, {Shopbell}, {Smol{\v c}i{\'c}}, {Taniguchi}, {Taylor},
  {White}, \& {Yan}}]{2013ApJS..206....3S}
{Scoville}, N., {Arnouts}, S., {Aussel}, H., {et~al.} 2013, \apjs, 206, 3

\bibitem[{{Scoville} {et~al.}(2007){Scoville}, {Aussel}, {Brusa}, {Capak},
  {Carollo}, {Elvis}, {Giavalisco}, {Guzzo}, {Hasinger}, {Impey}, {Kneib},
  {LeFevre}, {Lilly}, {Mobasher}, {Renzini}, {Rich}, {Sanders}, {Schinnerer},
  {Schminovich}, {Shopbell}, {Taniguchi}, \& {Tyson}}]{2007ApJS..172....1S}
{Scoville}, N., {Aussel}, H., {Brusa}, M., {et~al.} 2007, \apjs, 172, 1

\bibitem[{{Smol{\v c}i{\'c}} {et~al.}(2008){Smol{\v c}i{\'c}}, {Schinnerer},
  {Scodeggio}, {Franzetti}, {Aussel}, {Bondi}, {Brusa}, {Carilli}, {Capak},
  {Charlot}, {Ciliegi}, {Ilbert}, {Ivezi{\'c}}, {Jahnke}, {McCracken},
  {Obri{\'c}}, {Salvato}, {Sanders}, {Scoville}, {Trump}, {Tremonti}, {Tasca},
  {Walcher}, \& {Zamorani}}]{2008ApJS..177...14S}
{Smol{\v c}i{\'c}}, V., {Schinnerer}, E., {Scodeggio}, M., {et~al.} 2008,
  \apjs, 177, 14

\bibitem[{{Sutherland} \& {Saunders}(1992)}]{1992MNRAS.259..413S}
{Sutherland}, W. \& {Saunders}, W. 1992, \mnras, 259, 413

\bibitem[{{Wylezalek} {et~al.}(2013){Wylezalek}, {Galametz}, {Stern}, {Vernet},
  {De Breuck}, {Seymour}, {Brodwin}, {Eisenhardt}, {Gonzalez}, {Hatch},
  {Jarvis}, {Rettura}, {Stanford}, \& {Stevens}}]{2013ApJ...769...79W}
{Wylezalek}, D., {Galametz}, A., {Stern}, D., {et~al.} 2013, \apj, 769, 79

\bibitem[{{Zucca} {et~al.}(2009){Zucca}, {Bardelli}, {Bolzonella}, {Zamorani},
  {Ilbert}, {Pozzetti}, {Mignoli}, {Kova{\v c}}, {Lilly}, {Tresse}, {Tasca},
  {Cassata}, {Halliday}, {Vergani}, {Caputi}, {Carollo}, {Contini}, {Kneib},
  {Le F{\`e}vre}, {Mainieri}, {Renzini}, {Scodeggio}, {Bongiorno}, {Coppa},
  {Cucciati}, {de La Torre}, {de Ravel}, {Franzetti}, {Garilli}, {Iovino},
  {Kampczyk}, {Knobel}, {Lamareille}, {Le Borgne}, {Le Brun}, {Maier},
  {Pell{\`o}}, {Peng}, {Perez-Montero}, {Ricciardelli}, {Silverman}, {Tanaka},
  {Abbas}, {Bottini}, {Cappi}, {Cimatti}, {Guzzo}, {Koekemoer}, {Leauthaud},
  {Maccagni}, {Marinoni}, {McCracken}, {Memeo}, {Meneux}, {Moresco}, {Oesch},
  {Porciani}, {Scaramella}, {Arnouts}, {Aussel}, {Capak}, {Kartaltepe},
  {Salvato}, {Sanders}, {Scoville}, {Taniguchi}, \&
  {Thompson}}]{2009A&A...508.1217Z}
{Zucca}, E., {Bardelli}, S., {Bolzonella}, M., {et~al.} 2009, \aap, 508, 1217

\end{thebibliography}

\Online
\appendix
\section{Tables of host X-ray clusters and galaxy groups}
In this appendix the complete tables of candidate associations between radio AGNs found in this work and known clusters and groups catalogues are reported.

\begin{table*}
\caption{Candidate associations of radio AGNs and X-ray clusters from the catalogue of \citetads{2007ApJS..172..182F}. $RA_{cl}$, $dec_{cl}$ and $z_{cl}$ are the coordinates of the cluster centre. Column 4 is the distance between the cluster centre and the source of the AGN sample considered on the plane of the sky, while column 5 is the distance in redshift between the two. $L_X([0.1- 2.4] keV)$ and $\log(L_{1.4 GHz})$ are the X-ray and 1.4 GHz luminosities.}
\centering
\begin{tabular}{c c c c c c c}
\hline\hline
$RA_{cl}$ & $dec_{cl}$ & $z_{cl}$ & Distance (arcsec) & $\Delta z$ & $L_X([0.1- 2.4] keV)  (\frac{erg}{s \times cm^2})$ & $\log(\frac{L_{1.4 GHz}}{W\times Hz^{-1}})$ \\
\hline
$0.7 \le z < 1$ \\
\hline
150.41386 & 1.84759 & 0.969&	40.176    &0.015      &$2.9883\times 10^{43}$ &25.68  \\
150.27736 & 2.05303 & 0.908&	41.112    &-0.0521    &$9.6268\times 10^{42}$ &23.76 \\
150.02382 & 2.20323 & 0.942&	7.1856    &0.0988     &$1.9325\times 10^{43}$ &23.35 \\
149.64966 & 2.20925 & 0.954&	0.120744  & -0.0165   &$1.7111\times 10^{43}$ &24.34 \\
150.21454 & 2.28010 & 0.881&	26.3124   &-0.0419    &$1.3734\times 10^{43}$ &24.59 \\
149.95262 & 2.34188 & 0.942&	0.0144    &0.0156     &$1.5213\times 10^{43}$ &23.47 \\
149.95262 & 2.34188 & 0.942&	41.76     &-0.0002    &$1.5213\times 10^{43}$ &24.16 \\
149.92926 & 2.40902 & 0.874&	0.0072    & -0.0044   &$8.4425\times 10^{42}$ &23.58 \\
149.63988 & 2.34912 & 0.951&	41.112    & -0.0369   &$3.2975\times 10^{43}$ &24.06 \\
149.62355 & 2.39918 & 0.845&	0.108072  & 0.0051    &$1.9522\times 10^{43}$ &24.31 \\
150.15298 & 2.39447 & 0.899&	26.01     & 0.0193    &$9.4442\times 10^{42}$ &23.64 \\
149.66927 & 2.47365 & 0.957&	11.4408   & -0.0044   &$1.8343\times 10^{43}$ &23.92 \\
149.56174 & 2.42195 & 0.846&	0.0144    & 0.0332    &$1.7645\times 10^{43}$ &24.20 \\
150.00713 & 2.45343 & 0.731&	0.1476    & 0.0229    &$5.2795\times 10^{42}$ &24.15 \\
149.92343 & 2.52499 & 0.729&	30.8016   & 0.0016    &$1.1257\times 10^{44}$ &23.74 \\
150.10533 & 2.72392 & 0.727&	2.39364   & -0.0023   &$6.9659\times 10^{42}$ &23.89 \\
149.91772 & 2.70088 & 0.889&	3.00132   & -0.0355   &$2.1160\times 10^{43}$ &24.28 \\
150.58397 & 2.32155 & 0.720&	46.116    & 0.0182    &$4.2323\times 10^{42}$ &23.39 \\
150.05057 & 2.13923 & 0.959&	1.09008   & 0.1189    &$5.7184\times 10^{42}$ &24.19 \\
\hline
$1 \le z \le 2$ \\
\hline
150.76245 & 1.79362 & 1.258 & 10.4004 & -0.0051 & $5.5689\times 10^{43}$ & 26.43 \\
149.51855 & 2.09959 & 1.382 & 48.096 & 0.1839 & $7.8229\times 10^{43}$ & 24.00 \\
150.59309 & 2.53890 & 1.045 & 7.056 & -0.1063 & $3.1937\times 10^{43}$ & 23.63 \\
149.59763 & 2.44004 & 1.168 & 10.4076 & -0.023 & $2.6946\times 10^{43}$ & 26.86 \\
150.57024 & 2.49864 & 1.146 & 17.586 & -0.0083 & $2.5386\times 10^{43}$ & 24.13 \\
150.12646 & 1.99926 & 1.019 & 13.644 & -0.1744 & $1.2897\times 10^{43}$ & 23.65 \\
\hline
\end{tabular}
\label{Finoguenov}
\end{table*}

\begin{table*}
\caption{Candidate association between radio AGN and groups from the catalogue of \citetads{2012ApJ...753..121K}. $RA_{gr}$, $dec_{gr}$ and $z_{gr}$ are the coordinates of the group centre. $N_{spec}/N_{photo}$ is the number of sources with spectroscopic and photometric redshift respectively, column 5 and 6 are the distances from the group centre to the position of the radio AGN source on the plane of the sky and along the redshift direction respectively. $\log(L_{1.4 GHz})$ is the 1.4 GHz luminosity.}
\centering
\begin{tabular}{c c c c c c c}
\hline\hline
$RA_{gr}$ & $dec_{gr}$ & $N_{spec}/N_{photo}$ & $z_{gr}$ & Distance(arcsec) & $\Delta z$  & $\log(\frac{L_{1.4 GHz}}{W\times Hz^{-1}})$ \\
\hline
$0.7 \le z < 1$ \\
\hline
149.920567 & 2.521800 & 12/82 &  0.7297  & 17.6328  &  0.0023  & 23.74  \\
150.005475 & 2.451841 & 12/49 &  0.7311  & 8.1792   &  0.023   & 24.15  \\
150.212666 & 2.281762 & 7/14  &  0.8812  & 18.7812  & -0.0417  & 24.59  \\
149.552005 & 2.423054 & 5/25  &  0.8446  & 35.2692  &  0.0318  & 24.2   \\
149.914194 & 2.694681 & 5/54  &  0.8898  & 55.008   & -0.0669  & 24.05  \\
149.914194 & 2.694681 & 5/54  &  0.8898  & 28.5552  & -0.0347  & 24.28  \\
150.101070 & 2.268664 & 3/4   &  0.6855  & 47.16    & -0.0547  & 23.36  \\
150.094158 & 2.063360 & 3/7   &  0.725   & 53.064   &  0.01    & 23.61  \\
150.580055 & 2.328653 & 3/6   &  0.7274  & 28.3644  &  0.0256  & 23.39  \\
150.023066 & 2.516794 & 3/2   &  0.7473  & 31.554   &  0.0314  & 24.92  \\
150.508547 & 2.653296 & 3/7   &  0.8102  & 34.056   &  0.0103  & 25.02  \\
149.569460 & 2.418630 & 3/107 &  0.8529  & 30.258   &  0.0401  & 24.2   \\
150.515963 & 2.005564 & 3/11  &  0.8797  & 15.822   &  0.0089  & 24.29  \\
149.788439 & 2.757904 & 3/10  &  0.9093  & 24.6348  & -0.0272  & 24.16  \\
150.475201 & 1.626689 & 3/1   &  0.9713  & 32.6952  & -0.0236  & 23.64  \\
150.098295 & 2.056180 & 2/3   &  0.6375  & 47.34    & -0.0775  & 23.61  \\
150.017350 & 2.442392 & 2/5   &  0.6681  & 54.288   & -0.04    & 24.15  \\
149.725800 & 2.770682 & 2/7   &  0.7032  & 42.48    & -0.0508  & 23.98  \\
149.838840 & 1.683059 & 2/7   &  0.747   & 0.126828 &  0.0067  & 23.3   \\
149.731350 & 2.759319 & 2/6   &  0.7655  & 18.3024  &  0.0115  & 23.98  \\
150.231175 & 2.072644 & 2/1   &  0.7995  & 40.212   & -0.0594  & 23.61  \\
150.465750 & 2.423217 & 2/4   &  0.8266  & 9.9792   & -0.0006  & 24.4   \\
150.410760 & 1.801465 & 2/5   &  0.8456  & 13.6152  &  0.0372  & 24.77  \\
149.985860 & 2.325898 & 2/3   &  0.849   & 32.7384  & -0.0764  & 23.94  \\
150.091666 & 2.599807 & 2/4   &  0.8929  & 20.2572  & -0.0961  & 24.07  \\
\hline
$1 \le z \le 2$ \\
\hline
150.216932 & 2.273971 & 2/5   &  0.893   & 44.424   & -0.0299  & 24.59  \\
150.445160 & 1.845406 & 2/8   &  0.8943  & 20.6964  &  0.0232  & 24.6   \\
150.023318 & 2.205914 & 2/12  &  0.9409  & 5.7276   &  0.0977  & 23.35  \\
149.963878 & 2.363113 & 2/10  &  0.9454  & 50.04    &  0.0032  & 24.16  \\
149.649660 & 2.209250 & 2/7   &  0.9539  & 0.120744 & -0.0166  & 24.34  \\
149.843245 & 2.573333 & 2/8   &  0.9657  & 42.408   & -0.0285  & 25.08  \\
149.633185 & 2.457927 & 3/2   &  1.1688  & 40.716   & -0.0832  & 25.06  \\
150.128820 & 1.922544 & 2/1   &  1.0083  & 53.172   & -0.0832  & 23.64  \\
\hline
\end{tabular}
\label{knobel}
\end{table*}

\end{document}